\documentclass[fleqn,aps,prb,preprint,a4paper,onecolumn]{revtex4}

\usepackage{amsmath}
\usepackage{amssymb}
\usepackage[dvips]{graphicx}
\usepackage{color}
\usepackage{tabularx}
\usepackage{mathtools}
\usepackage{algpseudocode}
\usepackage{enumitem}
\usepackage{multirow}
\usepackage{epstopdf}  
\usepackage{bm}
\usepackage{color}
\usepackage{soul} 
\usepackage{makecell}
\usepackage[
bookmarks=true,
colorlinks,
linkcolor=blue,
urlcolor=blue,
citecolor=blue,
plainpages=false,
pdfpagelabels,
final,
breaklinks=true
]{hyperref}

\usepackage{threeparttable}

\definecolor{orange}{RGB}{255,127,0}

\newcommand{\XY}[1]{{#1}}

\newcommand{\recheck}[1]{{#1}}

\begin{document}

\title{A Tungsten Deep Neural-Network Potential for Simulating Mechanical Property Degradation Under Fusion Service Environment}
\author{Xiaoyang Wang}
\affiliation{Laboratory of Computational Physics,
  Institute of Applied Physics and Computational Mathematics, Huayuan Road 6, Beijing, P. R.~China}
\author{Yinan Wang}
\affiliation{School of Mathematical Sciences, Peking University, No.5 Yiheyuan Road Haidian District, Beijing, P.R.China 100871}
\author{Linfeng Zhang}
\affiliation{DP Technology, Beijing, 100080}
\affiliation{AI for Science Institute, Beijing, 100084}
\author{Fuzhi Dai}
\affiliation{DP Technology, Beijing, 100080}
\affiliation{AI for Science Institute, Beijing, 100084}
\author{Han Wang}
\email{wang\_han@iapcm.ac.cn}
\affiliation{Laboratory of Computational Physics, Institute of Applied Physics and Computational Mathematics, Fenghao East Road 2, Beijing 100094, P.R.~China}
\affiliation{HEDPS, CAPT, College of Engineering, Peking University, Beijing 100871, P.R.~China}

\begin{abstract}
Tungsten is a promising candidate material in fusion energy facilities. Molecular dynamics (MD) simulations reveal the atomistic scale mechanisms,
% \XY{of tungsten property degradation}, 
so they are crucial for the understanding of the macroscopic property deterioration of tungsten under harsh and complex service environment. The interatomic potential used in the MD simulations is required to accurately describe a wide spectrum of relevant defect properties, which is by far challenging to the existing interatomic potentials. In this paper, we propose a new three-body embedding descriptor and hybridize it into the Deep-Potential (DP) framework, an end-to-end deep learning interatomic potential model. 
Trained with the dataset generated by a concurrent learning method, the potential model for tungsten, named by DP-HYB, is able to accurately predict a wide range of properties \recheck{including elastic constants, stacking fault energy, the formation energies of free surfaces and point defects, which are included in the training dataset, and formation energies of grain boundaries and prismatic loops, the core structure of screw dislocation, the Peierls barrier and the transition path of the screw dislocation migration, which are not explicitly included in the training dataset. 
The DP-HYB is a good candidate for the atomistic simulations of tungsten property deterioration, especially those involving the mechanical property degradation under the harsh fusion service environment.
}
% Trained with the dataset generated by a concurrent learning method, the potential model for tungsten, named by DP-HYB, is able to accurately predict a wide range of properties including elastic constants, the formation energies of free surfaces, grain boundaries, point defects and defect clusters, stacking fault energy, the core structure of screw dislocation, the Peierls barrier and the transition path of the screw dislocation migration.
% Since most of the properties are not explicitly included in the training dataset, the strong generalizability of the DP-HYB model indicates that it is a good candidate for the atomistic simulations of tungsten property deterioration, especially those involving the mechanical property changing under the harsh service environment. 
\end{abstract}
\maketitle

\setlength{\parskip}{.5em}

\section{Introduction}
Tungsten (W) and W alloys are refractory, high strength materials that are proved to be promising candidate structural materials in fusion devices of ITER project\cite{zinkle2013multimodal}. 
Although W materials have multiple excellent properties such as high melting point, high thermal conductivity, high sputtering threshold and low atomic activity\cite{causey1999tritium,bolt2004materials,neu2007final,wittlich2009damage,rieth2013recent}, stable services under the extreme service environment in fusion devices are still challenging~\cite{hu2016irradiation,hasegawa2011property}. 
During the services, the W materials are subject to mutual influences of high dose rate of neutron irradiation, plasma flushing and high heat load, etc.~\cite{zinkle2009structural}, and the multi-physics coupling of these effects leads to severe deterioration of many properties critical to the stability of W materials.
Among all the properties of W, the degradation of its thermal-mechanical properties are of the central concern. 
For example, %W has a high ductile-brittle transition temperature (DBTT). 
% The irradiation damage worsens the ductility, which further weakens the mechanical stability of components in the fusion devices, thus increases the risks in safety. 
irradiation induced defects act as obstacles to the gliding dislocations, and play major roles in the degradation of ductility\cite{osetsky2021atomic}. \XY{The massive thermal load during W-plasma interaction leads to the formation of cracks beneath the W surface, and will degrade the mechanical strength\cite{ARAKCHEEV2015246}. Besides, the high service temperature induces the coarsening of grains or even re-crystallization, which raises ductile-brittle transition temperature~\cite{gietl2022neutron}.}
%Besides, the nucleation, \recheck{growth or propagation} of micro-cracks are also key processes to the brittleness.
All these processes under the \XY{ service environments} worsen
% \XY{\sout{the ductility and}}
the mechanical stability of the components in the fusion devices, thus increasing the risks to safety. 
To reveal how the processes occur requires the resolution of a series of critical atomistic events, which are not easily accessible to neither experimental approaches nor the continuum/meso scale simulations.
Therefore, the large-scale atomistic simulation serves as an important tool to study the underlying nano-micro scale mechanisms responsible for the deterioration of the mechanical properties of W.

\XY{The potential energy surface (PES) is of central importance in atomistic simulations.
To simulate property degradation under service conditions, the interatomic potential model is required to be applicable to a broad spectrum of properties. 
For example, a typical scenario under the multi-physics service condition is the poly-crystal W plastically deforming, driven by the external stress, while influenced by the irradiation induced damages. 
This case involves dislocation gliding, interaction with irradiation defects, pinning, de-pinning and interaction with grain boundaries.
To conduct reliable atomistic simulation in this case, the requirement for the potential is the accurate evaluation of properties, including elastic constants, dislocations, free surfaces, grain boundaries, point defects and their clusters by the same model.}

The PES can be modeled from the quantum mechanical principles like the density functional theory (DFT), which has been proved to achieve relatively high accuracy and reliability. 
DFT has long been the {\it de facto} standard for calculating the properties like lattice parameters, cohesive energy, elastic constants and the formation energy of point defects.
However, the computationally affordable sample size of DFT is usually limited to no more than $10^3$ atoms due to the typical $\mathcal O(N^3)$ computational complexity ($N$ being the number of atoms).
For the macro-scale observable properties, such as yield strength, plasticity and work hardening, which are linked directly to extended defects, one needs to simulate systems composed of millions of atoms, a scale that goes far beyond the capability of typical DFT calculations. Thus, to simulate these properties, the development of a computationally affordable PES is in demand.
The reliability of the interatomic potentials used in the complex service environment lies in two aspects\cite{zhang2019active}: (1) Good representability, \XY{i.e., the model can well reproduce all its training properties};
(2) Good generalizability, i.e., the model is able to describe \XY{properties not explicitly presented in the training datasets.} %two requirement for a potential(in general)
\recheck{The generalizability can be classified as in-distribution and out-of-distribution generalizabilities, which describe the ability of the model to interpolate within and extrapolate out-of the distribution of training data, respectively. 
For example, trained with liquid configuration, the in-distribution generalizability means the ability of the model to accurately predict on other liquid configurations that are not in the training data, while the out-of-distribution generalizability means the ability of the model to predict on solid configurations. 
}

The empirical or semi-empirical interatomic potential models the PES by a relatively simple analytical function form with tunable parameters, and is usually the method of choice in large scale atomistic simulation due to the $\mathcal O(N)$ computational complexity. Constant efforts have been made to develop reliable W interatomic potentials from decades ago, and up to now dozens of potentials for W have been proposed.
See, e.g., Refs.~\cite{finnis1984simple,ackland1987improved,johnson1989analytic,foiles1993interatomic,dai2007long,derlet2007multiscale,wang2013modified,marinica2013interatomic,mundim2001diffusion,juslin2013interatomic,wood2017quantumaccurate}. 
Most of the potentials use fixed formalisms such as Finnis-Sinclair (FS)\cite{finnis1984simple}, embedding atom method (EAM)\cite{ackland1987improved}  and modified embedded atom method (MEAM)\cite{baskes1992modified}. 
The parameters of a potential are tuned by fitting the prediction of the potential to the DFT calculation or a group of experimentally measured properties. 
The two approaches can also be combined to tune the potential model.

\XY{
The analytical function forms of the empirical interatomic potential are usually developed based on physical and/or chemical knowledge, thus the empirical models may have a strong ability for generalization, in both senses of in-distribution and out-of-distribution. 
A good example is the Marinica-13 potential~\cite{marinica2013interatomic}, which is trained by 6 point defects, 2 perfect body centered cubic (BCC) and face centered cubic (FCC)
and 10 liquid configurations, and is able to correctly predict many materials properties such as the formation energy of point defect clusters, the screw dislocation core-structure and a reasonably good Peierls barrier. 
The empirical interatomic potentials often suffer from representability issues, because the analytical function forms may not be flexible enough to fit a broad range of training targets.
As a consequence, the empirical potentials are usually fitted to a relatively small collection of targets, and it is not trivial to increase the training dataset without loss of accuracy. 

For the purpose of investigating the deterioration of the mechanical properties of W under service conditions, the state-of-the-art empirical potentials are not satisfactory. 
The Marinica-13~\cite{marinica2013interatomic} potential predicts wrong relative stability of $\langle111\rangle$ and $\langle100\rangle$ prismatic loops (see Fig.\ref{fig:loop}).
Two later potentials, proposed by Bonny et.~al.\cite{2017BonnyWRe} and Setyawan et.~al.\cite{2018WResetyawan}, improved from the EAM-2 and EAM-4 of Marinica-13, respectively, still meet the same challenge (see Fig.\ref{fig:loop}).  
The EAM potential proposed by Mason et.~al.\cite{Mason2017} is fitted to reproduce the interaction between vacancies in clusters and free surface formation energies, which have long been challenging for EAM potentials~\cite{Bonny_2014}.
This potential does not perform well when reproducing the generalized stacking fault energy ($\gamma$-line)~\cite{2020Evaluation}.
A more recent EAM potential proposed by Chen et.~al.~\cite{2018New}} is fitted to lattice parameters, elastic constants and formation energy of point defects. It agrees well with its DFT database on these properties, and can be generalizable to other properties such as the generalized stacking fault energy ($\gamma$-line), but it gives a qualitatively wrong prediction on the screw dislocation core structure (the \recheck{``Degenerate-core"}, see the discussion in Sec.~\ref{sec:screw-dislocation}). 
\XY{Unfortunately, $\gamma$-line, screw dislocation core structure and dislocation loop formation energy are key tungsten properties under service environment. Thus even using these most advanced empirical potentials, the simulation results concerning the irradiation induced mechanical property changes will be in question.
}

Recent development of Machine-Learning(ML) potential\cite{thompson2015spectral,shapeev2016moment,behler2007generalized,bartok2010gaussian,chmiela2017machine,schutt2017schnet,smith2017ani,han2017deep,zhang2018deep,zhang2018end,wood2017quantumaccurate} hold the promise of having both the accuracy of DFT and efficiency of empirical potentials\cite{2018Screw}. 
Previous successful applications of Gaussian Approximation Potentials (GAP)\cite{AP2012Gaussian} for W include two GAP potentials that considered different aspects of the properties. 
The GAP potential proposed by Szlachta, Bart\'ok and Cs\'anyi\cite{2014Accuracy} (GAP-1) considered properties of bulk, elastic, mono-vacancy, free surfaces, stacking fault energy and screw dislocation and excluded the self-interstitial atoms (SIA), 
while the GAP potential proposed by Byggm{\"a}star et.~al.~\cite{byggmastar2019machine} (GAP-2) additionally included SIAs, their clusters, vacancy clusters, short-range repulsion, liquid phase, distorted surfaces in the training dataset, but omitted the stacking fault energy and screw dislocation properties. 
The computational cost of the GAP potentials %are
is in proportional to the size of their training dataset, thus it is not common to include all relevant configurations in the training dataset for the sake of efficiency, and a balance between the accuracy and efficiency is carefully searched~\cite{2014Accuracy}.
This feature of GAP potentials makes it too expensive to take advantage of large databases, which further limits the ability of generalization to properties dissimilar to what are included in its training datasets. 
For example, the accuracy of GAP-2 on large SIA clusters would not be satisfactory until including di-SIA cluster configurations in the training dataset~\cite{byggmastar2019machine}, and without the explicit consideration of dislocation properties, the GAP-2 presents an obviously lower Peierls barrier than DFT\cite{VENTELON20133973}.
Thus, none of the GAP models can be directly used to conduct atomistic simulation on mechanical property degradation of W.
A more efficient implementation of ML potential on W is the spectral neighbor analysis potential (SNAP)\cite{wood2017quantumaccurate}. 
In this potential a subset of GAP-1 training database is utilized.
In addition, it included the configurations of equation of states, liquid phases and multiple vacancies in its training dataset. 
Meanwhile, it omitted the screw dislocation structures and generalized stacking fault. 
Thus extended defect properties such as generalized stacking fault energy and Peierls stress are overestimated by the SNAP\cite{SNAP-test} potential.
During the preparation of this manuscript, we noticed the publication of a new formalism of ML potential, namely, the quadratic-noise ML potential\cite{PhysRevMaterials.5.103803} for the crystal defect of W. This ML potentials takes a simpler form and targets mainly the properties of point-defect clusters and dislocations.
Although the robustness of generalization outperforms the GAP models due to its preconditioned linear-ML fitting, its representability is expected to be lower than the GAP formalism.

Benefited from the excellent ability of neural-networks to fit to high-dimensional, multi-variant functions, the deep-neural-network (DNN) based interatomic potentials have shown a great capability of describing multi-component materials' properties~\cite{zhang2019active,Jiang2020AccurateDP,zhang2021phase,wang2021generalized,marchand2020machine,bahramian2013study,urbanczyk2018neural,hu2019accurate,li2017cu,kobayashi2017neural,artrith2016implementation,artrith2012high}. 
In addition, the computational cost of DNN is determined by the size of neural-network and the size of the system, and is independent on the size of training datasets. This feature of DNN makes it practical to take advantage of relatively large datasets.
\recheck{Although the extensible and symmetries are strictly satisfied by the DNN potential models, the generalizability of them are often put in question. 
A well-known fact is that the out-of-distribution generalizability of DNNs is not expected~\cite{bartok2018machine}. 
The questions faced by DNN based potential models come to (1) how to generate a training dataset that covers the relevant configuration space as comprehensively as possible, and (2) how well can the model generalize in the sense of in-distribution, so all the desired properties can be reproduced at the DFT accuracy.
Since the out-of-distribution generalizability of DNNs should be entirely forgotten, in the rest of the paper, when we refer to the word ``generalizability", we mean the in-distribution generalizability.
}
% \XY{Although the flexible construct of DNN have no fixed form of physical basement, as long as the configuration space in the training database is explored comprehensively, the materials property will be well predicted based on the transferring from the knowledge in the database}. \XY{The validity of the extrapolations of the potential is dependent on the transferability of the DNN.}

In this paper, we use the Deep-Potential (DP) method to establish a highly generalizable interatomic potential for BCC W. This potential is highlighted by adopting our newly-designed symmetry-preserving descriptor, namely, three-body embedding, which, in addition to the bond-angle contributions encoded in the environment matrix, explicitly considers the bond-angle contributions in the embedding part of the descriptor~\cite{zhang2018end}. 
The formalism of the new descriptor is inspired by the nature of BCC transition metals~\cite{marinica2013interatomic}: their d-bands are not fully filled, which leads to a relatively complex shape of the electron density distribution with obvious angular characters~\cite{moriarty1990analytic}.
Thus, explicitly considering these bond-angle embedding  is expected to increase the representability of the model to transition metals such as W.
Trained with the large dataset \XY{automatically} generated by the concurrent learning strategy Deep-Potential GENerator (DP-GEN)~\cite{zhang2020dpgen}, the W DP model is proved to be in satisfactory agreement with DFT and/or experiment in a wide range of properties including bulk properties, %\DFZ{point defects and extended defects,}
elastic constants, formation energies of surfaces/interfaces, point  defects  and  their clusters  as  well  as  the  major  aspects  of  screw dislocation properties \XY{such as the core structure, the Peierls barrier and the dislocation trajectory during its migration.}
\XY{A majority of the investigated properties, i.e., the formation energy of prismatic loops and C15 Laves phase clusters, interaction energy between neighboring vacancies, screw dislocation core structures and migration path, Peierls barrier and grain boundary structures,} %\WH{Here we should explicitly say which properties are blind to the model}
% \XY{not explicitly presented in the DFT database of}
are  blind to the DP model in the sense that the corresponding configurations are not explicitly included in the training dataset. Thus, we argue that the DP model is of high generalizability.
Moreover, due to the high representability of DP models, this W DP potential can be feasibly adjusted to describe more properties of interest according to the purpose of different simulations.

The manuscript is organized as follows. In Section~\ref{sec:method}, we introduce the mathematical structure of the DP descriptor and the three-body embedding descriptor. 
Then we describe the details of the concurrent learning scheme. 
In Section~\ref{sec:Validation}, we present the benchmark results of DP models on properties of W, and make comparisons with results obtained by DFT calculations and by a wide range of previous empirical/ML potentials. Then we draw conclusive remarks in Section~\ref{sec:conclusion}.

\section{Method} \label{sec:method}

\subsection{The Deep Potential Model}
The DP model assumes that the potential energy of the system can be decomposed into atomic contributions $E = \sum_i E_i$, where $E$ and $E_i$ denote the system energy and the contribution due to atom $i$, respectively. 
The atomic energy $E_i$ depends on the position of atom $i$, and its near neighbors, whose distances to $i$ are less than the cutoff radius.
We denote the positions of atoms in the system by $\{\bm r_1, \cdots, \bm r_N\}$, and define the \emph{environment matrix} for $i$, which records all the relative positions of near neighbors to $i$, by
\begin{equation}
\label{eqn:envmat}
(\mathcal{R}_i)_{j,\cdot} = s(r_{ij})\times(\frac{x_{ij}}{r_{ij}},\frac{y_{ij}}{r_{ij}},\frac{z_{ij}}{r_{ij}}),
\end{equation}
where $r_{ij} = {\vert \bm r_{ij}\vert}$, $\bm r_{ij} = \bm r_j - \bm r_i$,  $(x_{ij}, y_{ij}, z_{ij})$ are the three Cartesian components of $\bm r_{ij}$.  $s(r_{ij})=f_c(r_{ij})/r_{ij}$ with $f_c$ being a switching function smoothly varies from 1 to 0 at the cutoff distance. 
The environment matrix has $N_m$ rows, where $N_m$ is the maximally possible number of neighbors, 
and Eq.~\eqref{eqn:envmat} presents the definition of the $j$-th row.
When the real number of neighbors, denoted by $N(i)$ is less than $N_m$, the environment matrix is padded with zeros. 
% From Eq.~\eqref{eqn:envmat}, we define the \emph{generalized environment matrix by}
The DP models the atomic energy contribution $E_i$ by
\begin{align}
    E_i = \mathcal F \Big(\mathcal D( \mathcal R_i ) \Big),
\end{align}
where $\mathcal D$ is the descriptor that maps the environment matrix to symmetry preserving features, and $\mathcal F$ is the fitting net that represents the energy dependency on the local atomic configuration. 
It is noted that the descriptor can be hybridized with several descriptors, i.e.~the concatenation of symmetry preserving features with different designs,
\begin{align}
    \mathcal D = \Big(
    \mathcal D^{(2)}, \mathcal D^{(3)}, \cdots 
    \Big).
\end{align}

The smooth edition of the DP~\cite{zhang2018end} 
proposes the following descriptor
\begin{align}\label{eqn:descriptor2}
    \mathcal D_i^{(2)} = 
    \frac{1}{N_m^2}
    ( \mathcal G_i^{(2),<} )^T 
    \tilde{\mathcal R}_i 
    (\tilde{\mathcal R}_i)^T
    \mathcal G_i^{(2)}
\end{align}
where $\tilde{\mathcal R}_i$ is the \emph{generalized environment matrix} defined by
\begin{equation}
\label{eqn:genvmat}
(\Tilde{\mathcal{R}}_i)_{j} = s(r_{i,j})\times(1, \frac{x_{ij}}{r_{ij}},\frac{y_{ij}}{r_{ij}},\frac{z_{ij}}{r_{ij}}),
\end{equation}
and $\mathcal G_i^{(2)}$ is the \emph{embedding matrix} involving two-atom distance and is defined as
\begin{equation}\label{eqn:embdmat2}
(\mathcal{G}_i^{(2)})_{j,\cdot} = \Big( G_1^{(2)}(s(r_{ij})), 
\cdots
G_{M_2}^{(2)}(s(r_{ij}))
\Big),
\end{equation}
where ${G}^{(2)}$ is the two-body embedding network, \XY{represented by a DNN mapping from a single value $s(r_{ij})$, through multiple hidden layers, to $M_2$ outputs (also the number of columns of $\mathcal G_i^{(2)}$).
% $M_2$ is the maximum width of the embedding network.
The Eq.~\eqref{eqn:embdmat2} defines the $j$-th row of the embedding matrix. 
The embedding matrix in total has $N_m$ rows and is pad with zeros if $N(i) < N_m$. 
The $\mathcal G_i^{(2),<} $ denotes a sub-matrix of $\mathcal G_i^{(2)} $ containing the first $M^<$ columns of $\mathcal G_i^{(2)} $.}
We refer the embedding matrix defined by Eq.~\eqref{eqn:embdmat2} as the \emph{two-body embedding matrix}, because it only involves distances between two atoms, i.e.~$r_{ij}$, in the construction. 
\recheck{More mathematical details regarding the two-body embedding network and embedding matrix are found in the Supplementary Material Sects.~IIA and IIB.}

In most cases, the two-body descriptor is expressive enough to achieve a satisfactory accuracy, even in the most stringent test of the accuracy of potential energy such as the phase diagram of water~\cite{zhang2021phase}.
However, as a typical BCC transition metal, W has unfilled $d$-bands thus has more complex shape of electronic density distribution. In these metals, the bond-angle contribution are essential to the proper description of materials properties\cite{moriarty1990analytic}.
This makes the bond-angle contribution in W playing significant roles in the prediction of potential energy\cite{marinica2013interatomic}. 
Though in the descriptor of the smooth edition of DP~\eqref{eqn:descriptor2}, the bond-angle contribution has been considered by the multiplication of the generalized environment matrices $\tilde{\mathcal R}_i(\tilde{\mathcal R}_i)^T$, further incorporating the bond-angle contribution explicitly into the embedding matrix is expected to increase the representability for BCC transition metals.

This inspires us to propose the following \emph{three-body embedding tensor}
\begin{align}\label{eqn:embdmat3}
(\mathcal{G}_i^{(3)})_{jk,\cdot} &= 
\Big(
G_1^{(3)} \big( (\theta_i)_{jk} \big),
\cdots,
G_{M_3}^{(3)} \big( (\theta_i)_{jk} \big)
\Big),
\end{align}
\recheck{which is an order 3 tensor with the first two indices running over all neighbors and the third one goes from $1$ to $M_3$. 
The vector $G^{(3)}$ of length $M_3$ is represented by a DNN and is trainable. 
In Eq.~\eqref{eqn:embdmat3},}
\begin{align}\label{eqn:thetai}
(\theta_i)_{jk} &\equiv 
    ({\mathcal{R}}_i)_{j,\cdot}\cdot{(\mathcal{R}}_i)_{k,\cdot}
    =
    s(r_{ij})s(r_{ik})\frac{\bm r_{ij}\cdot\bm r_{ik}}{r_{ij} r_{ik}},
\end{align}
% which is an order 3 tensor considering the angle formed by two neighbors $j$ and $k$ of atom $i$ by the dot product of row $j$ and row $k$ of the environment matrix ${\mathcal{R}}_i$.
\recheck{is the dot product of row $j$ and row $k$ of the environment matrix ${\mathcal{R}}_i$, and encodes the bond-angle information of the two neighbors $j$ and $k$ of atom $i$.}
We thus define a new descriptor $\mathcal{D}_i^{(3)}$ from the three-body embedding tensor as
\begin{align}\label{eqn:descriptor3}
\mathcal{D}_i^{(3)} &= \frac{1}{N_m^2}\theta_i :\mathcal{G}_i^{(3)} 
= \frac{1}{N_m^2} \sum_{jk=1}^{N_m} (\theta_i)_{jk} (\mathcal{G}_i^{(3)})_{jk}, 
\end{align}
% where $:$ denote the double contraction of the $j$ and $k$ indices. 
\recheck{where $:$ denote the double summation over the $j$ and $k$ indices of the matrix $\theta_i$ (defined by Eq.~\eqref{eqn:thetai}) and the 
three-body embedding tensor $\mathcal{G}_i^{(3)}$.}
It is straightforward to prove that the descriptor Eq.~\eqref{eqn:descriptor3} is invariant under translational, rotational and permutational transforms.
\XY{In the Supplementary Materials Sects.~IIC and IID, we provide all detailed information of the mathematical formula of the three-body embedding tensor and descriptor, and provide the proof of the symmetries of the descriptor.
% s, the structures of embedding/fitting neural networks, the training procedures including hyper-parameters, and the proof of the symmetries of the descriptors.
}

The computational cost of the descriptor $\mathcal D^{(2)}$ is proportional to the number of neighbors in the neighbor list, while the complexity of the $\mathcal D^{(3)}$ is in proportional to the square of the number of neighbors. 
On the other hand, the precise description of the neighbor angles are only critical for the neighboring atoms within the first few neighbor shells~\cite{zhang2018deep}, 
and for the neighbors far away embedding the distance is enough for describing the local configuration. 
Therefore, we adopt a relatively small cutoff radius for the three-body embedding (4~\AA) while a relatively large cutoff (6~\AA) for the two-body embedding, 
then hybridize the two descriptors as $\mathcal D = (\mathcal D^{(2)}, \mathcal D^{(3)})$. 
In this work we refer to the DP using this hybridized descriptor as DP-HYB. 
The newly proposed DP approach is extensively compared with the smooth edition of DP (denoted by DP-SE2) that uses the descriptor $\mathcal D = \mathcal D^{(2)}$.

\subsection{\label{sec:DPGEN}DP-GEN Scheme}
We use the concurrent learning strategy named DP-GEN~\cite{zhang2019active,zhang2020dpgen} to generate the optimal training dataset in the sense that it is the most compact and adequate dataset to guarantee a uniform accuracy of DP in the relevant configuration space.
DP-GEN is a close-loop iterative workflow, including exploration, labeling and training steps.
Starting from an initial dataset, an ensemble of DP models is trained.
Then one of the models is used to explore the configuration space by simulation techniques like molecular dynamics (MD), Monte Carlo (MC), structure optimization or enhanced sampling.
The DP prediction error on the explored configurations is estimated by the deviation of the predictions of the ensemble of DP models, and only a small subset of the configurations with large errors are selected for labeling, i.e.~for the DFT calculations of energy, forces and virial tensor.
The iteration is converged when the prediction error on all explored configurations are tolerable.
The DP-GEN scheme has successful applications in the Al-Mg binary~\cite{zhang2019active} and  Al-Mg-Cu ternary alloys\cite{Jiang2020AccurateDP}, water in a very large thermodynamic region~\cite{zhang2021phase}, Ag-Au nanoalloys~\cite{wang2021generalized}, etc.

\paragraph{Initial dataset.} The initial dataset is composed of three parts:
\begin{enumerate}
\item The equilibriated \XY{unit cells} of the \XY{BCC}, the \XY{FCC}, the hexagonal close-packed (\XY{HCP}) and diamond structures.
\item Artificially strained and perturbed structures.
Hydrostatic strain is applied to equilibriated structures by changing lattice parameters ranging from 96\% to 106\% at a step 2\%.
Perturbations are then exerted on each structure under strain.
The atom positions are randomly perturbed with a maximal displacement of 0.01~\AA.
The cell is randomly perturbed by $3\%$.
\item AIMD trajectory of the deformed structures.
Setting each compressed structure under perturbation as initial configuration, AIMD simulations are conducted with 5 steps under temperature 100K.
\end{enumerate}
All the data are labeled with DFT calculations (see part \emph{c.~Labeling} for more details).
In other words, the DFT-calculated energy, atomic forces and virial tensor of each structure are recorded.  %\XY{The initial dataset make it possible for the models trained during the DP-GEN iterations stably sample the configuration space of BCC W.}

\paragraph{Exploration.}
The \texttt{LAMMPS} package\cite{plimpton1995lammps} compiled with the \texttt{DeePMD-kit}\cite{wang2018deepmd} support is employed to perform Deep Potential molecular dynamics (DPMD)\cite{zhang2018deep} simulations for the exploration of the configuration space. 
\XY{The exploration uses 4 subsets of configurations as the initial configurations for the MD sampling.
The initial configurations, simulation ensemble, temperatures and pressures conditions are summarized in the following and in the Table~\ref{tab:datasets}.
\begin{enumerate}
    \item The strained and perturbed $2\times2\times2$ supercell BCC W bulk. NPT ensemble. Temperature 50 to 5100~K. Pressure $-2$ to 5~GPa. 
    \item The strained and perturbed $3\times3\times3$ supercell BCC W bulk. NPT ensemble. Temperature 50 to 5100~K. Pressure $-2$ to 5~GPa. 
    \item (111),(110) and (112) free surfaces. NVT ensemble. Temperature 300 to 1800K.
    \item Locally perturbed $2\times2\times2$ and $3\times3\times3$ supercells of BCC W bulk. NVE ensemble.
\end{enumerate}
}

During the exploration, the deviation of force predictions of four DP models, trained with identical hyper-parameters but different random seeds, is used to estimate the error in the force prediction. If the maximal deviation of atomic forces is higher than 0.20~eV/\AA\ but lower than 0.35~eV/\AA, the configuration is \XY{considered as a candidate configuration} and sent for labeling. 
% \XY{\sout{Delete: The overall dataset of of the DP models consists of more than 40000~ frames of configurations.}}

\paragraph{Labeling.}
The labels of the candidate configurations, i.e.~the energy, force, and virial tensor, are computed by DFT with exchange-correlation modeled by the generalized gradient approximation (GGA) proposed by Perdew, Burke and Ernzerhof (PBE)~\cite{Perdew1996PBE}. 
The DFT calculations were conducted using VASP\cite{kresse1996efficient,kresse1996efficiency} package. The Brillouin zone is sampled by the Monkhorst-Pack method with a grid spacing of 0.16\AA$^{-1}$. The projector-augmented-wave (PAW) method is used and the energy cut-off of the plane-wave basis set is set to 600~eV. \XY{The $6s$ and $5d$ electrons are considered valence electrons.} The convergence criterion for the self-consistent field iteration is set to $10^{-6}$~eV. 
The same DFT parameters are also used for labeling the initial dataset. 

\paragraph{Training.}
In each iteration, four models are trained simultaneously using the same dataset and  hyper-parameters, with the only difference being the random seeds employed to initialize the model parameters. 
The sizes of the hidden layers of the two-body embedding
nets $G^{(2)}$ are (20, 40, 80), while the three-body embedding $G^{(3)}$ has hidden layers of sizes
(4, 8, 16). The hidden layers of the fitting nets are set to (240, 240, 240).
The Adam stochastic gradient descent method\cite{Kingma2015adam} with the default hyper-parameter settings provided by the \texttt{TensorFlow} package~\cite{abadi2015tensorflow} is used to train the DP models.
The learning rate is exponentially decayed with starting and final learning rates set to $ 1\times10^{-3} $ and $ 5\times10^{-8} $, respectively.
In each DP-GEN iteration the DP model is trained with $ 4\times10^{5
}$ steps.
After the DP-GEN iterations converge, the productive models are trained with $ 2.4\times10^{7}$ steps.
\recheck{The size of the DNNs used in the DP models, and other hyper-parameters are provided in the Supplementary Material Table SI.}

\paragraph{Refinement.}
The productive DP models are {firstly} refined with DFT labeled structures of SIA structures, the $\gamma$-line and (100) surface structure. 
SIA data is generated via additional DP-GEN iterations, with three types of SIA structures (namely $\langle111\rangle,\langle110\rangle$ and $\langle100\rangle$ dumbbells) as initial configurations, the explored temperature ranges from 50K to 600K.
$\gamma$-line structures are obtained directly from the DFT $\gamma$-line calculation along $\langle111\rangle$ directions on $\{110\}$ and $\{112\}$ planes.
We also included relaxed and unrelaxed (100) free surface data in the refining dataset.
The refined models are trained for $1\times10^{6}$ steps with model parameters initialized by the productive model.
% During the model refining trains,
The starting learning rate is set to $ 1\times10^{-4} $, and \recheck{the starting prefactors of energy, force and virial tensors, $p^{start}_\epsilon$,$p^{start}_f$,$p^{start}_\xi$ are set to 1.0, 1.0 and 0.9, respectively (See the Supplementary Material for the definition of the prefactors). The} 
other hyper-parameters are the same as those used to train the productive model.
\recheck{Then, the DP models are further refined by the isolated W atom energy.
All the hyper-parameters are the same with those used in the first refinement except that the number of training steps is set to $4\times 10^6$.}
\recheck{The energy of an isolated atom is calculated with spin-polarized DFT.}
% The energy of isolated atom is calculated with DFT using the same parameters with those used to label other data, except that the spin-polarization calculation is performed
Benchmark results are based on the model after the refinements.

 \XY{The composition of the training dataset is provided in Table~\ref{tab:datasets}. 
The training database consists of a total number of 43,648 configurations (1,103,542 local atomistic environments). 
The root mean square error (RMSE) of the energy (normalized by the number of atoms), the atomic forces, and the virial tensor (normalized by number of atoms) on the entire training dataset are $6.958\times 10 ^{-3}$~eV/atom, $1.278\times 10 ^{-1}$~eV/\AA\ and $1.186\times 10 ^{-1}$~eV/atom, respectively. The training error on each subset of the training dataset is shown in Table \ref{tab:datasets}.
}
%\resizebox{\textwidth}{!}{
\begin{center}
\begin{table}

\caption{\XY{The components of the training dataset. 
The initial dataset, and the sub-datasets generated by DP-GEN MD explorations starting from different initial configurations under various thermodynamic conditions are summarized. The bulk samples are heated up to 5100K to guarantee the coverage of liquid structures.
We also list the number of configurations ($n_{\textrm{conf}}$), number of the local atomistic environments ($n_{\textrm{env}}$) and training errors of energies and forces in the sense of RMSE of each sub-dataset.
}}\label{tab:datasets}
\begin{threeparttable} 
{\scriptsize
\begin{tabular}{c|ccccc}
\hline
\hline
Initial confs./Initial data & Ensemble & $n_{\textrm{conf}}$ & $n_{\textrm{env}}$ & RMSE$_E$(eV/Atom) & RMSE$_F$(eV/\AA)\\% & RMSE$_V(eV)$ \\
\hline
Initial dataset & - & 7060 & 17720& $4.815\times 10^{-3}$ &$7.711\times 10^{-2}$\\% &$2.477\times 10^{-2}$\\
\hline
2x2x2 Bulk BCC W & NPT 50-5100K -2 to 5Gpa & 4379& 70064& $1.186\times 10^{-2}$ &$2.478\times 10^{-1}$\\% &$7.201\times 10^{-2}$\\
\hline
3x3x3 Bulk BCC W & NPT 50-5100K -2 to 5Gpa & 2395& 129330& $7.508\times 10^{-3}$  &$2.601\times 10^{-1}$\\%&$3.773\times 10^{-2}$\\
\hline
(110),(112) Free Surface & NVT 300-1800K & 5477& 49770& $1.210\times 10^{-2}$ &$1.575\times 10^{-1}$\\% &$7.802\times 10^{-2}$  \\
\hline
(111) Free Surface & NVT 300-1800K & 1536& 146240& $1.711\times 10^{-3}$ &$1.136\times 10^{-2}$\\% &$1.505\times 10^{-2}$  \\
\hline
Local Perturbation & NVE & 19943& 500120& $4.295\times 10^{-3}$ &$8.676\times 10^{-2}$\\% &$2.839\times 10^{-2}$  \\
\hline
\hline
Refine Dataset & Ensemble & $n_{\textrm{conf}}$& $n_{\textrm{env}}$ & RMSE$_E$(eV/Atom) &   RMSE$_F$(eV/\AA) \\% & $RMSE_V(eV)$ \\
\hline
(100) Free Surface & NVT 300-1800K & 1315 &52600 &$2.290\times 10^{-3}$ &$1.180\times 10^{-1}$\\% &$8.459\times 10^{-3}$ \\
\hline
Generalized Stacking Fault& - & 286&  4752&$8.564\times 10^{-3}$ &$5.690\times 10^{-2}$\\% &$9.665\times 10^{-1}$ \\
\hline
Point defects& NVT 50-600K & 1257& 132946& $3.980\times 10^{-4}$ &$3.348\times 10^{-2}$\\% &$3.322\times 10^{-3}$\\
\hline
Isolated Atom& - & 1& 1& $7.004\times 10^{-3}$ &0\\% &$3.322\times 10^{-3}$\\
\hline
\hline
\end{tabular}
}
\end{threeparttable}
\end{table}
\end{center}

\section{Validation of the Deep Potentials}\label{sec:Validation}
In this section, benchmark results of DP models of W are presented.
The presented results include bulk properties, formation energies of point defects and prismatic loops, screw dislocation properties, generalized stacking fault energies, formation energies of free surfaces and typical grain boundaries. These properties are critical to the mechanical properties of W under service environment.
Our goal is to test if the W DP-HYB and DP-SE2 models are able to accurately predict all these properties.
Comparisons are made among the DP-HYB, DP-SE2, DFT and other interatomic potentials including the empirical and machine learning potential models.

Due to the large amount of empirical interatomic potentials reported in literature, it is impossible to comprehensively test all of them, so we only consider a limited subset, i.e.~the potentials by Ackland and Thetford et.~al.~(AT)~\cite{1987AcklandAT}, Juslin and Wirth (JW)~\cite{juslin2013interatomic}, Mason et.~al.~(MN)~\cite{Mason2017}, Chen and Li et.~al.~(CL)~\cite{2018New}. 
Two EAM potentials, MV-2-B by Bonny et.~al.~\cite{2017BonnyWRe} and MV-4-S by Setyawan et.~al.~\cite{2018WResetyawan}, modified from the potentials MV-2 and MV-4 developed by Marinica et.~al.~\cite{marinica2013interatomic}, are also considered. 
For AT, JW and MN, we mainly take the reference values from previous literature\cite{1987AcklandAT,juslin2013interatomic,Mason2017}, and for CL, MV-2-B and MV-4-S we reproduce the properties by ourselves for comparison.
Two GAP machine learning potentials are also included for comparisons, i.e.~the GAP model by Csanyi et.~al.~(GAP-1)~\cite{2014Accuracy},
and by Byggm\"asta et.~al.~(GAP-2)~\cite{byggmastar2019machine}. 
GAP-1 was publicly available, but is currently not available from the broken (\url{http://www.libatoms.org/} accessed on  Mar.30 2022) and redirected (\url{https://libatoms.github.io/GAP/}  accessed on  Mar.30 2022) web-links. 
% Currently GAP-1 is inaccessible. (At date Oct.4 2021, we cannot access GAP-1 on {http://www.libatoms.org/ and https://libatoms.github.io/GAP/}) .
Thus we mainly take the presented results from the paper of GAP-1~\cite{2014Accuracy}. For GAP-2, we take the presented results from Ref.\cite{byggmastar2019machine} and calculate the un-presented properties using the published GAP-2 potential. 

\subsection{Bulk Properties}
The lattice parameter ($a_0$), cohesive energy of BCC/FCC lattices ($E_{\textrm{BCC}}$,$E_{\textrm{FCC}}$) and three independent elastic constants ($C_{11}$, $C_{12}$ and $C_{44}$) of BCC W are reported in Table \ref{tab:Basic}.
% DP predicted results are compared with our DFT calculations, EAM potentials, GAP potentials and previous DFT/experimental data. 
\recheck{DP predicted results are compared with our DFT calculations, while the EAM potentials and the GAP potentials are compared with the corresponding fitting targets.}

Accurate prediction of these ground-state properties is the basis of evaluating all other surface and defect properties.  The W DP models have satisfactory agreement with our DFT calculation on the predictions of lattice parameters, cohesive energies and elastic constants, since these properties are well-presented in the initial training dataset.
The DP-HYB model has a slightly higher accuracy than DP-SE2 model on cohesive energies due to the improved representability. 
\recheck{The ground-state properties are the fitting targets of the investigated empirical potentials, and all the potentials can accurately reproduce the values in the training datasets.}
\recheck{When comparing across different EAM potentials, they show notable differences in the lattice parameters, because they are fitted to different target values.}
% The EAM potentials show notable differences in predictions of lattice constant, since they are fitted to different target values. 
For example, CL potential was fitted to the experimental value 3.165~\AA,  MV-2-B and MV-4-S agree with the DFT calculation that gave~3.14~\AA~\cite{2000Symmetrical}.
However, due to the strict requirement of self-consistence in training dataset, DP-SE2 and DP-HYB models are fitted only to our DFT data, which are, like other DFT results~\cite{2000Symmetrical,byggmastar2019machine,2010Helona}, slightly different from the experimental values.
Likewise, as the BCC crystal of W is sampled in a wide range of stress during the DP-GEN iterations, the elastic constants of both DP models are in good agreement with the DFT, which gives a higher $C_{11}$ and lower $C_{44}$ than experimental results\cite{simmons1965single} due to the pseudopotential used in the calculation\cite{PhysRevLett.79.2073}. 

\begin{center}
\begin{table}
\caption{Lattice parameters, cohesive energies and elastic constants predicted by EAM, GAP and DP models and our DFT calculation.
\recheck{The experimental values are provided in parenthesis.}
% Both DP models have satisfying predictions on lattice constant and ground-state cohesive energy of BCC lattice.
$\rm a_{0,BCC}$ and $\rm a_{0,FCC}$ denote the lattice parameters of BCC and FCC W at zero temperature, respectively. $\rm E_{BCC}$ and  $\rm E_{FCC}$ denote the BCC and FCC lattice cohesive energies. %calculated by the differences between energy per atom in the BCC/FCC lattice and the DFT calculated energy of an isolated atom.
$\rm C_{11}, C_{12}, C_{44}$ are independent elastic constant components of the BCC crystal.
% Empirical/GAP potentials can well-describe these properties as well. 
\XY{AT, JW and MN are fitted to experimental data. The CL and MV EAM potentials and the machine learning potentials are compared with their fitting targets.} 
% \WH{Only FCC phase??}
% \XY{Note that the differences among potentials on these basic properties are mainly caused by the different databases (or fitting targets). }
\XY{The * in the table indicates that the property is included in the training data as a fitting target, but the value is not presented in the reference.}
}\label{tab:Basic}

{
%\resizebox{\textwidth}{!}{
\begin{tabular}{c|ccccccc}
\hline
\hline
Potential&$\rm a_{0,BCC}$(\AA) &$\rm E_{BCC}(eV)   $ &$\rm C_{11}(GPa)$ &$\rm C_{12}(GPa)$ &$\rm C_{44}(GPa)$ &$\rm a_{0,FCC}$(\AA) &$\rm E_{FCC}(eV)   $ \\
\hline
MV-2-B\cite{2017BonnyWRe} &3.14&-8.9&522&203&160&3.793&-8.53\\
MV-4-S\cite{2018WResetyawan} &3.143&-8.9&522&202&161&3.738&-8.62\\
MV fitting target\cite{marinica2013interatomic}& (3.165)\cite{landolt1991numerical} & -8.9\cite{landolt1991numerical}
&523\cite{landolt1991numerical}&203\cite{landolt1991numerical}
&160\cite{landolt1991numerical}&4.054\cite{2010Generalized}&-8.43\cite{2010Generalized}\\
AT\cite{1987AcklandAT}&3.165&-8.9&522&204&161&3.927&-8.75\\
JW\cite{juslin2013interatomic} &3.165&-8.9&522&204&161&3.927&-8.75\\
MN\cite{Mason2017} &3.164&-8.9&526&204&161&3.927&-8.75\\
Experimental&(3.165)\cite{Kittel2005}&(-8.9)\cite{Kittel2005}&(522)\cite{simmons1965single}&(204)\cite{simmons1965single}&(161)\cite{simmons1965single}&-&-\\
CL\cite{2018New} &3.165&-8.9&523&204&161&4.103&-8.62\\
CL fitting target\cite{2018New} &3.165&-8.9&523&204&161&4.00&-8.33\\
GAP-1\cite{2014Accuracy} &3.181& -     &518&198&143& - & - \\
GAP-1 fitting target\cite{2014Accuracy}&3.181&-&517&198&142&-&-\\
GAP-2\cite{byggmastar2019machine} &3.185&-8.39&526&200&149& * & * \\
GAP-2 fitting target\cite{byggmastar2019machine}&3.185&-8.39&522&195&148&*&*\\
DP-SE2 &3.172&-8.46&{526}&192&138&4.023&-7.96\\
DP-HYB &3.172&-8.47&{543}&203&141&4.023&-7.97\\
DFT (this work) &3.171&-8.47&{548}&200&147&4.023&-7.98\\
\hline
\hline
\end{tabular}
}
%}
\end{table}
\end{center}

\begin{figure}
\begin{center}
\includegraphics[width=0.98\textwidth]{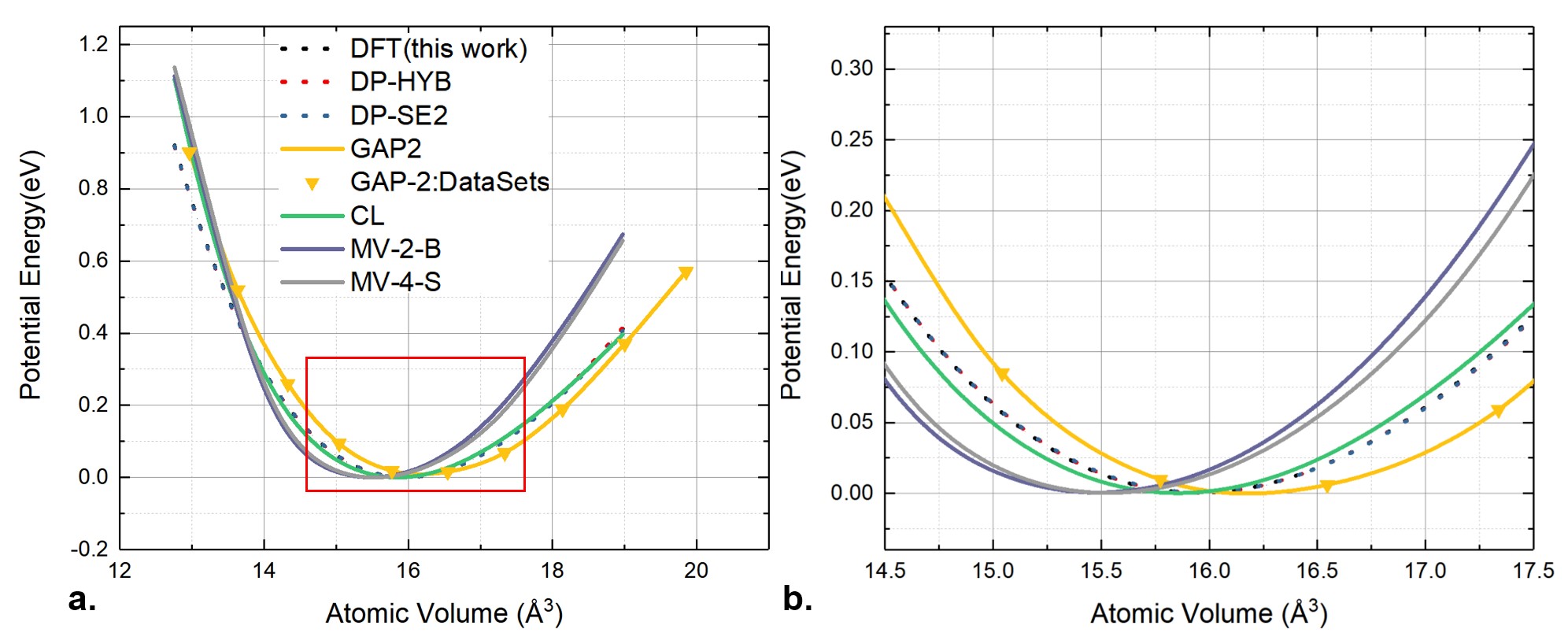}\\[5pt]  % insert figure
\caption{EOSs of BCC W predicted by DFT, DP models and three previous EAM models. 
All the curves are shifted by the energy of the stable BCC structure. a) EOS curve in atomsitic volume ranging from 12~\AA$^3$ to 19~\AA$^3$. b) Enlarged figure near bottom of the EOS curves. }
\label{fig:eos}
\end{center}
\end{figure}

The equations of states (EOS) of BCC W predicted by the DP models are shown in Fig.~\ref{fig:eos}.
We calculated the EOSs of EAM potentials CL, MV-2-B, MV-4-S,
% \XY{\sout{and plotted them  for comparison}
\XY{and take the EOS calculation result of GAP-2 from Ref.~\cite{byggmastar2019machine}}. 
All EOSs are shifted by \recheck{their corresponding ground-state energies of the BCC W}. 
Both DP-SE2 and DP-HYB are in almost perfect agreement with our DFT calculation.
The main difference between the EOSs calculated by the EAM and DP models are attributed to the different predictions of the lattice parameter.
As the system is compressed (volume smaller than the equilibrium volume), the increment of energy of all EAM potentials is faster than the DFT calculation. 
A faster increment of energy of MV-2-B and MV-4-S is also observed as the system is expanded (larger volume side).
GAP-2 exhibits a similar tendency of change of EOS against volume with both DP and DFT results. GAP-2 predicts the minimum EOS at around ~16.2\AA$^3$, slightly larger than the results predicted by DP models, due to the different pseudopotential used when constructing the database~\cite{byggmastar2019machine}. 

Both the DP-SE2 and DP-HYB models show good agreements with DFT calculation on the basic bulk properties of BCC W. Many other defect structure properties depend on these basic properties. To further validate the representability and generalizability of W DP models, in the following subsections the surface and defective structure properties are investigated. 

\subsection{Free Surfaces}

\begin{center}
\begin{table}
\caption{\XY{Formation energies of free surfaces with different miller indices are} calculated using DP models and DFT, and previous results using EAM and GAP models are also presented for comparison. 
\recheck{The EAM potentials AT, JW, MV-2-B, MV-4-S and CL are not fitting against the surface formation energies. 
The EAM potential MN, machine learning potentials GAP-1, GAP-2, DP-SE2 and DP-HYB have surface training data. Their fitting targets are presented along with the corresponding models.}
Data in the {parenthesis} is the unrelaxed surface energy. The units of free surface formation energy is: $J/m^2$.}\label{tab:Surface}

\begin{threeparttable}
\begin{tabular}{c|cccccc}
 
\hline
\hline
Potential &$(111)$ &$(122)$ &$(110)$ &$(112)$ &$(120)$ &$(100)$ \\
\hline
AT\cite{1987AcklandAT} &3.300&-&2.571&3.045&-&2.923\\
JW\cite{juslin2013interatomic} &3.300&-&2.575&3.045&-&2.923\\
MV-2-B\cite{2017BonnyWRe} &2.963&-&2.306&2.752&-&2.723\\
MV-4-S\cite{2018WResetyawan} &3.217&-&2.508&2.99&-&2.93\\
CL\cite{2018New} &3.263&-&2.541&2.989&-&2.893\\
MN\cite{Mason2017} &4.155&-&3.497&3.866&-&3.841\\
Fitting Target of MN&4.453&-&4.005&4.181&-&4.646\\

GAP-1\cite{2014Accuracy} &3.557&-&3.268&3.461&-&4.037\\
GAP-2\cite{byggmastar2019machine} &3.525&3.589&3.348&3.428&3.781&4.021\\
Fitting Target of GAP&3.556&-&3.268&3.460&-&4.021\\
DP-SE2 &\makecell[c]{3.543\\(4.040)}&\makecell[c]{3.682\\(4.050)}&\makecell[c]{3.366\\(3.418)}&\makecell[c]{3.420\\(3.720)}&\makecell[c]{3.581\\(3.874)}&\makecell[c]{3.937\\(4.295)}\\
DP-HYB &\makecell[c]{3.540\\(4.051)}&\makecell[c]{3.625\\(4.002)}&\makecell[c]{3.333\\(3.402)}&\makecell[c]{3.433\\(3.759)}&\makecell[c]{3.647\\(3.899)}&\makecell[c]{3.923\\(4.308)}\\

DFT (this work)&\makecell[c]{3.534\\(4.053)}&\makecell[c]{3.618\\(3.941)}&\makecell[c]{3.315 \\(3.380)}&\makecell[c]{3.453\\(3.797)}&\makecell[c]{3.689\\(4.005)}&\makecell[c]{4.048\\ (4.342)}\\
\hline
\hline
\end{tabular}
\end{threeparttable}
\end{table}
\end{center}

The formation energies of free surfaces with different miller indices are calculated using the DP models and DFT, and are shown in Table.~\ref{tab:Surface}.
% The predicted formation energy of free surfaces with different miller indexes are shown in Table \ref{tab:Surface}.
The DP models predict that the surface with the lowest formation energy is (110), in agreement with the DFT calculation and the reference.
% In agreement with DFT values and reference values, the surface with the lowest formation energy is (110).
The DP models achieve high accuracy in the (111), (110), (112) and (100) surfaces.
Noticing that these surface structures present in the training dataset, the high accuracy indicates good representability of the DP models.
% Since the DP models are trained with the (111), (110) and (112) surface structures and refined with the (100) surface structures, they achieve high accuracy in these surfaces.
Meanwhile, the DP models are accurate in describing the (122) and (120) surfaces, whose structures are not explicitly included in the training data.
% \XY{\sout{,which implies a satisfactory ability of generalization}}
\recheck{Noticing that these surfaces can decompose into the free surfaces in the training dataset, thus the accuracy is a consequence of generalization (in the sense of in-distribution).}
% \XY{However, these surfaces can decompose into the free surfaces included in the datasets. Thus, it is straightforward that DP models may also give reasonable evaluations of the formation energies of these free surfaces.}
Moreover, the DP models are able to correctly describe the un-relaxed free surfaces (presented in parentheses in Table.~\ref{tab:Surface}) of the investigated miller indices.
The accurate evaluation of free-surface formation energy can also be achieved by GAP models. By contrast, the accuracy of EAM potentials is lower than the DP models and the GAP potentials. Since the free surface formation energies are usually not the fitting target of EAMs, \XY{except the MN potential.  }
 
The accurate prediction on free surface properties ensures the reliability of the DP models in simulations of surface-related processes, such as cleavage fracture~\cite{2014Fracture}, ad-atom surface diffusion~\cite{Jansson_2020}, surface self-assembly~\cite{2017NanoCone} and nano-indentation\cite{2016nanoindentation}.

\subsection{Grain Boundaries}
Grain Boundaries (GBs) are interfaces between differently oriented grains. GBs play dominant roles in many observable properties of W poly-crystals, including textures, radiation resistance and overall mechanical responses to external strain.
GB structures are not explicitly presented in the DP training dataset, thus the accuracy of GB formation energy prediction by DP models is mainly guaranteed by the generalizability  of the models.

\begin{figure}
\begin{center}
\includegraphics[width=0.9\textwidth]{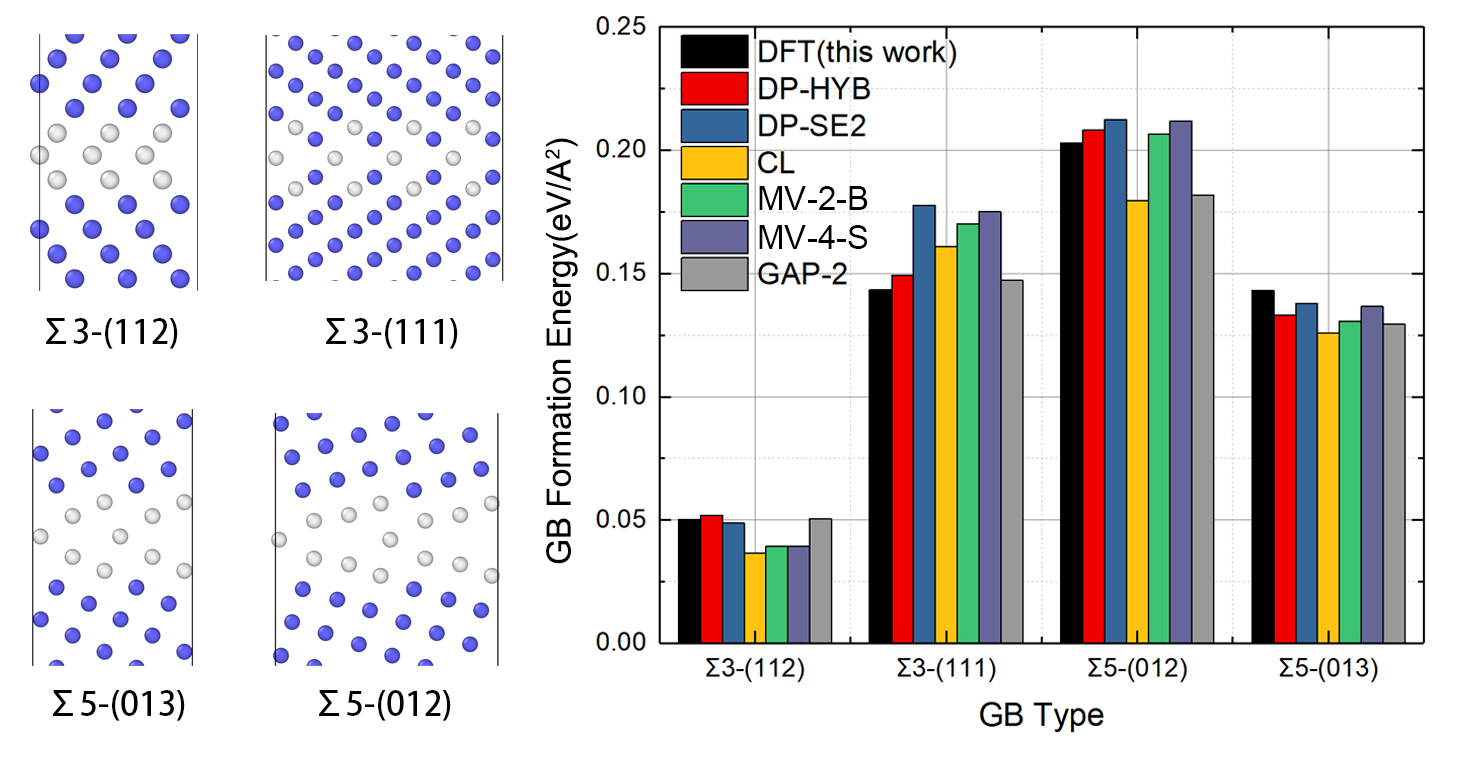}\\[10pt]  % insert figure
\caption{GB structures and formation energies predicted by DP models in comparison with DFT value (in this work) and other interatomic potential calculated values. Atoms are colored according to the results of common neighbor analysis: atoms in BCC lattice are colored blue and the atoms near the GB is colored gray.
}\label{fig:GB}
\end{center}

\end{figure}

The calculated GB formation energies of the DP-HYB, DP-SE2, EAM potentials, GAP-2 and DFT are presented in Fig.\ref{fig:GB}.
The calculated GBs are among the most frequently investigated symmetric tilt GBs: $\Sigma 3$-$(111)$, $\Sigma 5$-$(012)$, $\Sigma 5$-$(013)$ and $\Sigma 3$-$(112)$.
DP-HYB shows good agreement with DFT on all GBs presented in Fig.\ref{fig:GB}.
The accuracy of the DP-SE2 is comparable to DP-HYB in most cases, except the $\Sigma 3$-$(111)$ GB, in which DP-SE2 presents a larger deviation from DFT than all other potentials.
The DP-HYB is more accurate than all other EAMs on the formation energies of $\Sigma 3$-$(111)$ GB and $\Sigma 3$-$(112)$ GB. For $\Sigma 5$ GBs, all potentials show good agreement with DFT, thus DP-HYB has no obvious advantages. Although not considering GBs in their database, GAP-2 also have a good prediction on the GB formation energies.

\subsection{Point Defects and Prismatic Loops}
The formation energy of SIAs and a vacancy are shown in Table~\ref{tab:Point defect}.
Three dumbbell (db) SIA configurations, i.e.~$\langle 111 \rangle$, $\langle 110 \rangle$ and $\langle 100 \rangle$, are investigated.
The DP-SE2 presents that the formation energy of $\langle 111 \rangle$ dumbbell is only 0.09~eV lower than that of $\langle 110 \rangle$ dumbbell, which does not agree with DFT calculations showing that the $\langle 111 \rangle$ dumbbell is roughly 0.3~eV more stable than the $\langle 110 \rangle$ dumbbell. The disagreement should be attributed to the limited representability of only using two-body embedding in the descriptor, since the accuracy of DP-SE2 cannot be improved by using larger networks, more training data nor longer training procedure.
DP-HYB, by contrast, correctly predicts the formation energy difference between the dumbbell structures, we believe it is due to the better resolution of the bond-angle contribution in the atomic local environment in SIA structures, making it possible to distinguish different SIA structures. The point-defect formation energies of W are the main fitting targets of EAM potentials and GAP-2. Thus, although maintaining a satisfactory accuracy, DP-HYB has no obvious advantage on this property over the existing potentials.

\begin{center}
\begin{table}
\caption{Formation energy of differently oriented interstitial dumbbells and mono-vacancy predicted by EAM potentials, GAP models, and DP models. \XY{(units: eV)}}\label{tab:Point defect}
\resizebox{\textwidth}{!}{
\begin{threeparttable}
\begin{tabular}{c|cccccccccccc}
\hline
\hline
Type & AT\cite{1987AcklandAT} & JW\cite{juslin2013interatomic} & MV-2-B\cite{2017BonnyWRe} & MV-4-S\cite{2018WResetyawan} & MN\cite{Mason2017} & CL\cite{2018New} & GAP-1\cite{2014Accuracy} & GAP-2\cite{byggmastar2019machine} & DP-SE2 & DP-HYB & DFT\tiny{(this work)} & Reference	\\
\hline
$\langle111\rangle$ db &8.92&9.52&10.4&10.4&8.97&9.46&-&10.39&9.68&9.82&9.91&9.55\cite{2006Self},10.44\cite{2012Ab}\\
$\langle110\rangle$ db &9.62&10.18&10.86&10.89&9.68&9.8&-&10.60&9.77&10.17&10.25&9.84\cite{2006Self},10.82\cite{2012Ab}\\
$\langle100\rangle$ db &9.8&10.3&10.46&12.81&9.8&11.01&-&12.11&11.45&11.70&11.96&11.49\cite{2006Self},12.87\cite{2012Ab}\\
Vacancy &3.63&3.63&3.49&3.82&3.73&3.54&3.29&3.32&3.84& 3.30 &3.24&3.56\cite{2006Self},3.24\cite{PhysRevMaterials.5.103803}\\
\hline
\hline
\end{tabular}
\end{threeparttable}}
\end{table}
\end{center}

\XY{
% Interaction between vacancies in neighboring positions is a peculiar feature of W. 
The interaction between the first-nearest-neighbor (1NN) vacancies is reported to be  repulsive (with binding energy, $E_b(\textrm{1NN}) = -0.1$~eV~\cite{BECQUART200723}) or weakly attractive ($E_b(\textrm{1NN}) =0.048$~eV~\cite{Mason2017}), but that of the second-nearest-neighbor (2NN) vacancies is shown to be strongly repulsive ($E_b(\textrm{2NN})=-0.35$~eV~\cite{Heinola_2017}, $-0.286$~eV~\cite{Mason2017}). 
%This feature can be qualitatively reproduced by MN, GAP-2 and DP-HYB. 
By DP-HYB, $E_b$ between the 1NN vacancies is {$0.069$}~eV, and between 2NN vacancies the $E_b$ is {$-0.303$}~eV.
The di-vacancy configurations are not explicitly included in the training dataset of DP-HYB.
GAP-2 explicitly includes the 1NN and 2NN di-vacancies in their training database and can accurately reproduce the DFT calculation~\cite{byggmastar2019machine}.
Evaluating the 1NN and 2NN vacancy interactions have long been a challenge to the empirical potentials. 
Among them, the MN potential predicts qualitatively correct binding energies as $E_b(\textrm{1NN})=0.17$~eV and $E_b(\textrm{2NN})=-0.13~$eV\cite{Mason2017}. 
% , thus it is qualitatively correct in evaluating the interactions between vacancies.
}

Although point defect properties are important indicators of potential quality, individual point defects are hardly observable, and \XY{the impact of individual point defects} to the materials mechanical properties are relatively weak.
Instead, their clusters, formed due to the agglomeration of the point defects, may substantially alter the mechanical properties via their interactions with dislocations.
Here, we primarily consider two major forms of SIA clusters: prismatic loops and C15-Laves phase clusters.

\begin{figure}
\begin{center}
\includegraphics[width=0.9\textwidth]{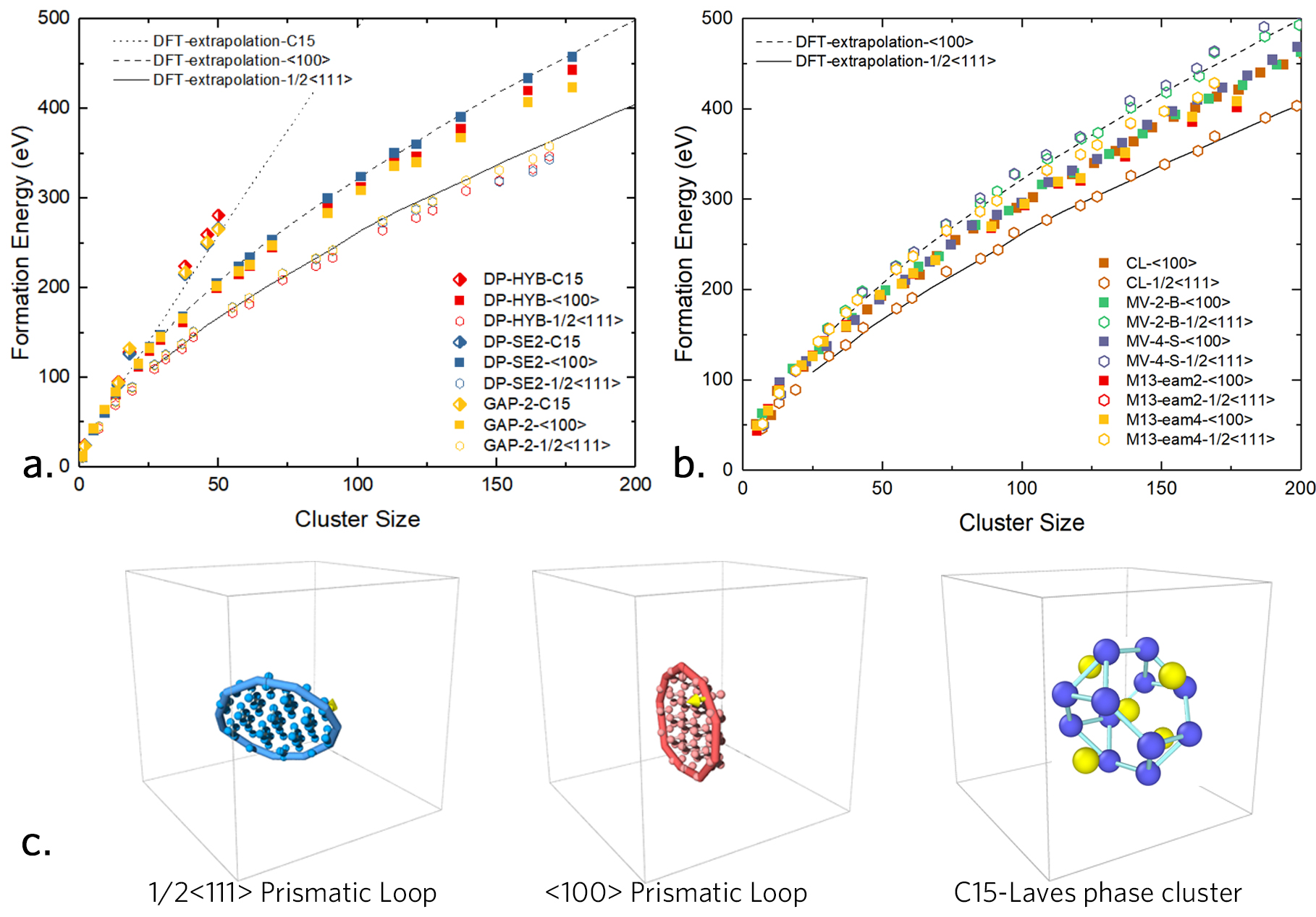}\\[10pt]  % insert figure
\caption{Prediction of formation energies of  $1/2\langle111\rangle$, $\langle100\rangle$ prismatic loops and C15-Laves phase cluster by DP-HYB model. Comparison is made among the DP models, DFT extrapolation\cite{2016Ab} and \XY{EAM potentials.} a) Formation energy of $1/2\langle111\rangle$, $\langle100\rangle$ loops and  C15-Laves phase clusters predicted by DP-HYB, DP-SE2 and GAP-2 potential. b) Formation energy of $1/2\langle111\rangle$ and $\langle100\rangle$ loops predicted by EAM potentials. \XY{Note that MV-2-B and MV-4-S potentials are modified from Marinica-13 (M13) EAM-2 and EAM-4, respectively. The predictions of 1/2$\langle111\rangle$ loops by the M13-EAM-2 and M13-EAM-4 models almost overlap with each other}. c) The structures of $1/2\langle111\rangle$, $\langle100\rangle$ loops and  C15-Laves phase clusters in BCC W. }\label{fig:loop}
\end{center}

\end{figure}

Prismatic loops are the most common configurations of SIA clusters, and are frequently observed in experiments~\cite{ELATWANI2018206}.
Prismatic loops in BCC metals can be categorized into two types according to their Burgers vectors, as shown in Fig.\ref{fig:loop}c, the 1/2$\langle 111 \rangle$ loop and the $\langle 100 \rangle$ loop~\cite{2013Solving}, with the latter playing a more important role in radiation induced hardening at service temperatures due to its lower diffusivity~\cite{reza2020thermal}.
Fig.~\ref{fig:loop}a shows that the formation energies of $\langle 111 \rangle$ and $\langle 100 \rangle$ prismatic loops predicted by the DP-HYB model are \recheck{in good agreement with} the DFT extrapolation by a discrete-continuum(DC) model~\cite{2016Ab}.
It is noted that large prismatic loops cannot be calculated directly via DFT, thus are not presented in the training data.
The accurate prediction of the formation energies demonstrates the DP-HYB's generalizability  to large defect clusters.

Liu et.~al.~pointed out that most EAM potentials are not able to accurately evaluate the relative stability of large loops~\cite{2020Evaluation}.
We plot the loop formation energies predicted by CL, MV-2-B and MV-4-S in Fig.~\ref{fig:loop}b for comparisons.
All the EAM models have similar predictions of formation energies of the $\langle 100 \rangle$ loops.
The MV-2-B and MV-4-S potentials predict that the 1/2$\langle 111 \rangle$ loop is less stable than $\langle 100 \rangle$, which is contradictory to the DFT extrapolation. 
The AT potential (not shown in Fig.~\ref{fig:loop}) gives nearly equal formation energies of the 1/2$\langle 111 \rangle$ and the $\langle 100 \rangle$ loops, JW and MN EAM potentials (not shown in Fig.~\ref{fig:loop}) may predict the correct relative stability of the two loops, but the difference between the formation energies of loops are too small to distinguish the two types of loops\cite{2020Evaluation}. 
The CL potential, as the only exception of the EAM potentials, gives qualitatively correct formation energy of the $\langle111\rangle$ loop and the correct relative stability, but the accuracy of the $\langle100\rangle$ loop formation energy is lower than that of the DP-HYB model. 
GAP-2 potential benefits from the enhanced generalizability due to the consideration of SIA-clusters, thus may also well predict the relative stability of the $\langle111\rangle$ loop and $\langle100\rangle$ loop. The formation energies of either prismatic loops or individual self-interstitial atoms is out of the concern in the database the GAP-1, \XY{nor did its reference paper present loop formation energies.} 

In addition, we used the DP models to predict the formation energy of a C15-Laves phase self-interstitial cluster\XY{, whose structure is not explicitly presented in the training dataset.
The C15-Laves phase cluster} is a unique type of SIA cluster other than the prismatic loops, and is able to transform into either type of the prismatic loops\cite{zhang2015formation}.
Unlike the case in BCC Fe\cite{PhysRevMaterials.5.103803}, the C15 cluster in BCC W is never the stablest configuration at any cluster size.
This is predicted by both the DP model and the DFT extrapolation, see Fig.~\ref{fig:loop}. 
C15 Laves phase cluster can also be correctly predicted by GAP-2, since the C15 structure is explicitly presented in the database of GAP-2.

The SIA clusters, such as the two-dimensional prismatic loops and the relative ordered three-dimensional C15-Laves phase clusters, are blind to the DP-HYB model, thus the ability of accurately predicting the formation energies of the clusters is mainly attributed to the generalizability  of the DP-HYB model and the training dataset.
The correct prediction of formation energy proves the DP-HYB model to be reliable candidate potential for simulating the SIA defect cluster evolution, which is critical to the post-irradiation micro-structure. 

\subsection{Generalized Stacking Faults}
\begin{figure}
\begin{center}
\includegraphics[width=0.98\textwidth]{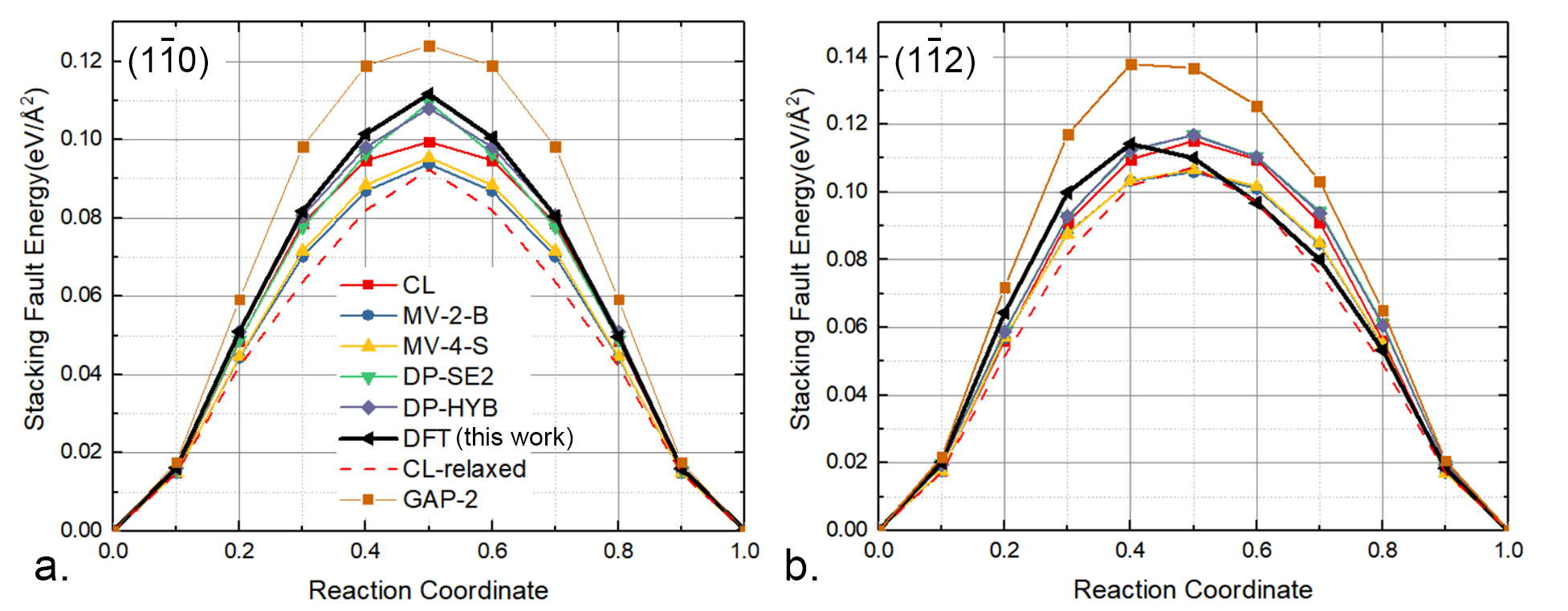}\\[5pt]  % insert figure
\caption{Prediction of generalized stacking fault energy along $\langle111\rangle$ direction on a) (1$\bar1$0) and b) (1$\bar1$2) plane. }\label{fig:gamma}
\end{center}
\end{figure}

The generalized stacking fault (GSF) energy landscape is defined as the variation of energy on displacing one part of the crystal against the other on a specific plane ($\gamma$ plane).
The GSF energy has strong implications on the possible slip systems of a crystal~\cite{2010Effect} and the existence of meta-stable stacking faults. GSF energy along a specific direction is referred to as $\gamma$-line.
We calculate the GSF energies by using the DP models and other EAM/GAP models along the $\langle 111 \rangle$ direction of the (1$\bar1$2) $\gamma$ plane and (1$\bar1$0) $\gamma$  plane, and compare the $\gamma$-lines with those calculated by DFT in Fig.~\ref{fig:gamma}.
In order to reproduce the GSF energy of CL potential reported by Chen et.~al.~\cite{2018New}, we have to use the un-relaxed GSF energies, while for all the other potential models and the DFT calculation we use the relaxed GSF energies.
The relaxed $\gamma$-lines of CL potential are also plotted for comparison.

The $\gamma$-lines predicted by both DP models are in good agreement with DFT.
The EAM predictions of the $\gamma$-lines are somehow lower than DFT at the (1$\bar1$0) $\gamma$  plane and the (1$\bar1$2) $\gamma$  plane.
The $\gamma$-lines are also well presented in the training datasets of GAP-1 model, but are omitted by the GAP-2 model. According to ref.\cite{2014Accuracy}, GAP-1 may well-predict the $\gamma$-line, but we observe a significant overestimation of $\gamma$-line by GAP-2 in Fig.~\ref{fig:gamma}.

\subsection{Screw Dislocation}
\label{sec:screw-dislocation}
In the BCC metals, dislocations are the main carriers of plastic deformation~\cite{PO2016123}. Plastic deformation in W is dominated by the slow-moving screw dislocation at up to modest temperature\cite{2018Screw}.
The high migration barrier leads to the low mobility of the screw dislocation, which further determines many key features of the mechanical properties~\cite{Samolyuk_2012}. For example, in pure W the screw dislocation migration takes a thermally activated "kink-pair" formation mechanism~\cite{2012Quantum} with an activation energy of 1.7-2.1~eV\cite{marinica2013interatomic}, leading to low mobility at room temperature. This critical feature is determined by the core structure of screw dislocation, which may also significantly affect the possible slip planes\cite{Bonny_2014}. Therefore, for the purpose of investigating the mechanical properties of W, an accurate description of the core structure is the fundamental requirement for the interatomic potential.

\begin{figure}
\begin{center}
\includegraphics[width=0.95\textwidth]{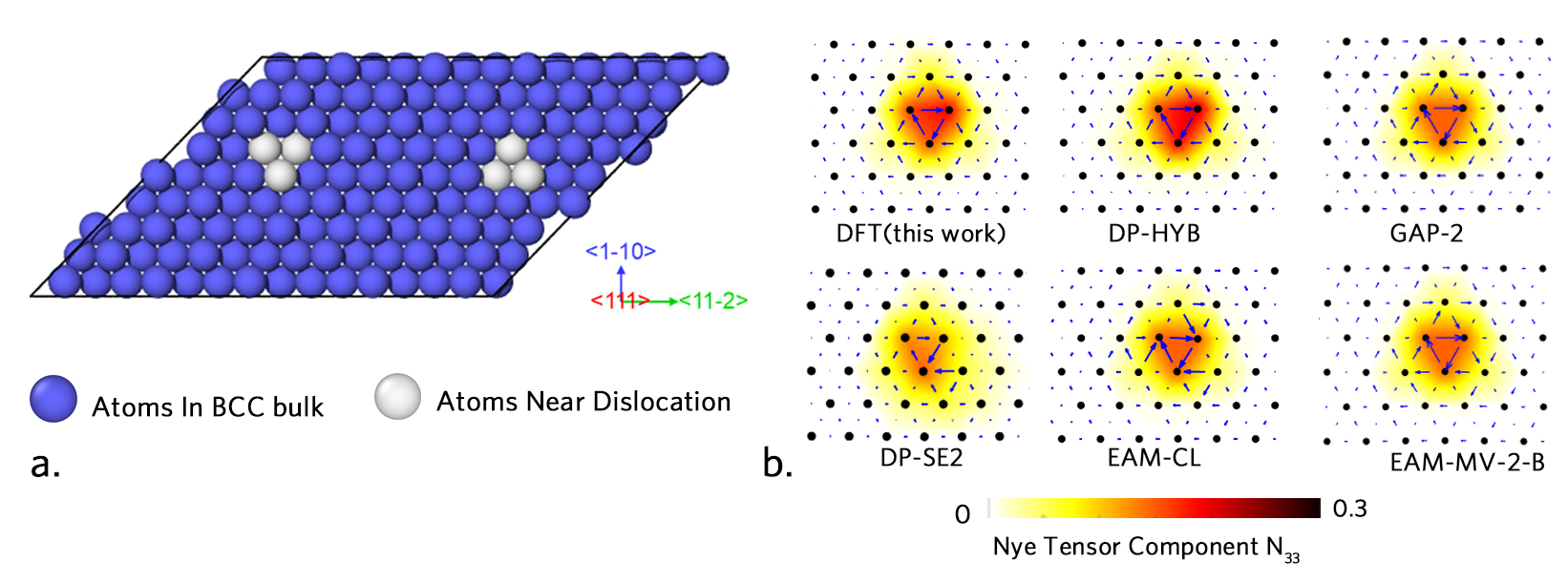}\\[5pt]  % insert figure
\caption{ a) Schematics illustration for the dipole arrangement used for calculating the core-structure of screw dislocation in DFT simulations. b) Differential displacement vector and nye-tensor distribution for ground-state screw dislocation core structure predicted by DFT, DP-HYB, DP-SE2 , CL, MV-2-B and GAP-2 potentials.} \label{fig:screw_statics}
\end{center}
\end{figure}

A schematic illustration of the system we used to calculate the core structure of the screw dislocation using DFT is presented in Fig.~\ref{fig:screw_statics}~(a). The system is composed of 135 atoms with 2 screw dislocations aligned in opposite directions in a dipole arrangement~\cite{2014Accuracy}. The dislocation Burgers vectors are $[111]$, and $[\bar{1}\bar{1}\bar{1}]$, respectively.
The dipole arrangement ensures the periodic boundary condition applied to all dimensions. The same configuration is used to calculate the Peierls barriers with interatomic potentials. These screw dislocation structures are not included in the training database of the DP models.

In Fig.~\ref{fig:screw_statics}~(b), the differential displacement map (DD-map) and Nye tensor distribution near the screw dislocation cores obtained by the DFT, DP-HYB, DP-SE2, CL/MV-2-B EAM potentials and GAP-2 are presented.
The DP-HYB model predicts that the screw dislocation in BCC W exhibits a non-planar, non-degenerate core structure with three-fold symmetry, and the dislocation core does not spread towards three nearby $\langle112\rangle$ directions. The dislocation core structure of the DP-HYB model is in satisfactory agreement with our DFT calculation and other DFT results reported in literature~\cite{Cereceda_2013,VENTELON20133973,2012The}.
The MV-2-B and MV-4-S potentials are able to predict correct core structure as well~\cite{2017BonnyWRe,2018WResetyawan}.
The GAP-1 potential explicitly includes the dislocation structures in its training dataset, thus being able to predict the correct core structure.
The GAP-2 potential does not explicitly consider the screw dislocation structures in their training dataset, and is able to predict the correct screw core structure.
By contrast, DP-SE2 predicts a split-core structure, and the CL EAM potential gives a hard-core structure, which are inconsistent with the DFT calculations.

\begin{figure}
\begin{center}
\includegraphics[width=1.00\textwidth]{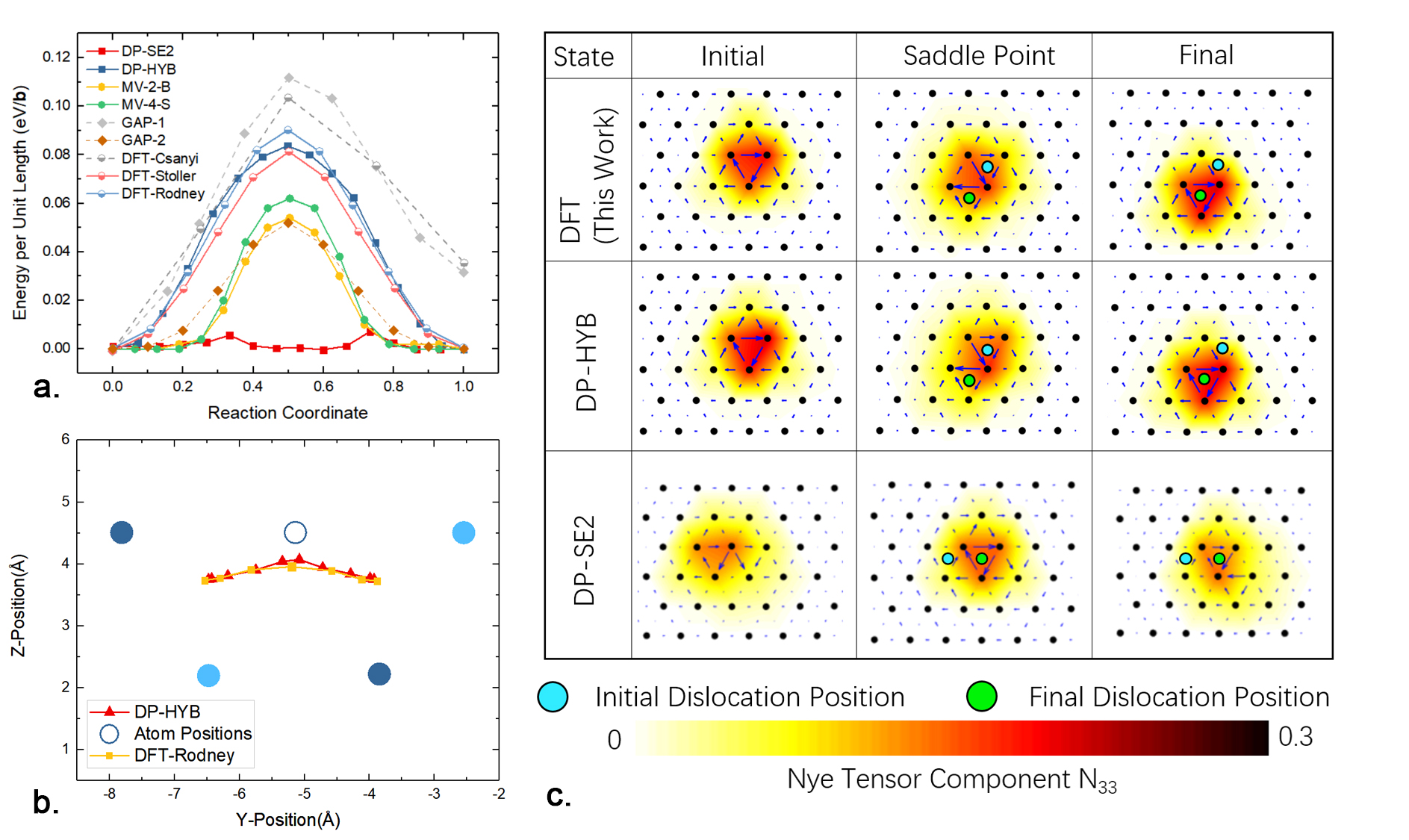}\\[5pt]  % insert figure
\caption{ a) Peierls barrier obtained by DFT, DP-HYB and DP-SE2 models, EAM models, GAP models and reference DFT values, respectively. We also take DFT calculated Peierls barriers by Rodney et al\cite{VENTELON20133973}, Csanyi\cite{2014Accuracy} et al. and Stoller\cite{2012The} et al as references. DP-HYB can predict the correct Peierls barrier value {84~meV/b}, which is close to DFT calculated results, but the DP-SE2 model fails to reproduce Peierls barrier value. 
b) Screw dislocation trajectory along its MEP predicted by DP-HYB model and previous DFT\cite{VENTELON20133973}. c) Dislocation core structure DD-maps along their transition path predicted by DFT and DP models. DP-HYB can predict the correct ground-state dislocation core structure, and well reproduces the core structure at saddle point.} \label{fig:screw_dynamics}
\end{center}
\end{figure}

The intrinsic migration barrier per unit length of screw dislocation, named the Peierls barrier, determines the mobility of screw dislocation.
We calculate the minimum-energy-path (MEP) of screw dislocation migration by the nudged-elastic-band (NEB)\cite{2000A} method, and present the MEPs in Fig.~\ref{fig:screw_dynamics}~(a).
The DP-HYB model gives a Peierls barrier of {84~meV/b}, which is in good agreement with the DFT calculations ranging from 70meV/{b} to 105meV/{b}~\cite{marinica2013interatomic,2014Accuracy,VENTELON20133973} and our DFT calculation result {89~meV/b}.
By contrast, the prediction of DP-SE2 on Peierls barrier is drastically lower than the reference DFT values.
We attribute the unphysical prediction of DP-SE2 to the lack of representability, because the DP-SE2 could not be improved by explicitly including the configurations along the MEP path in the training dataset.
This drawback is significantly improved by the three-body embedding formalism employed in the DP-HYB model.
The MV-2-B and MV-4-S EAM potentials predict Peierls barriers of {54 and 62}~meV/b, respectively.
The GAP-1 potential also has a good prediction on the Peierls barrier~\cite{2014Accuracy}, since it considered the screw dislocation structure in the dataset.
The GAP-2 potential, however, omitted the screw dislocation properties for the purpose of keeping the dataset relatively small to reuse in other BCC metals~\cite{byggmastar2019machine} (see ref.\cite{byggmastar2020gaussian} for GAP potentials for V, Mo and Ta et.~al.), thus predicted a lower Peierls barrier than GAP-1 and the DFT references.

The dislocation core structures of the DFT, DP-HYB and DP-SE2 at the initial, saddle and final states of the migration path are plotted and compared in Fig.~\ref{fig:screw_dynamics}~(c).
DP-HYB is able to predict the correct ground-state three-fold core structure, and the split-core structure at the saddle point. DP-SE2 disagrees with DFT on core-structure at the saddle point.

A special feature of BCC metals is the break down of Schmid-Law, namely the yield stress of BCC metals under mechanical loading depends not only on the orientation of the shear plane, resulting in the so-called twining/anti-twining (T/AT) asymmetry\cite{kraych2019non}.
This asymmetry is physically connected to the departure of the dislocation migration MEP away from the straight path connecting two nearby non-degenerate core positions\cite{2016Plastic}.
To test whether the non-schmid effect can be predict by the DP-HYB model, we extracted the dislocation core trajectory along the MEP via the cost-function method~\cite{VENTELON20133973,dezerald2014ab}.
The trajectory of the dislocation core calculated by DP-HYB is close the DFT calculation~\cite{kraych2019non}, as is observed from Fig.~\ref{fig:screw_dynamics}~(b).
This indicates that the non-Schmid effect can be correctly exhibited in the simulations using the DP-HYB model.
Since the DP-SE2 does not present the reasonable core structure and Peierls barrier, the dislocation trajectory calculation for DP-SE2 is not presented.
Again, it should be emphasized no screw dislocation structure is presented in the training database of the DP-HYB model,
thus the accurate prediction of these properties is mainly attributed to the generalizability  of the model.
The agreement with DFT on screw dislocation core structure, Peierls barrier, and dislocation trajectory, proves that the DP-HYB model is reliable for simulating the screw dislocation-dominated plastic deformation under mechanical load.

\subsection{Efficiency}
In order to benchmark the efficiency of the ML potentials, we conduct  MD simulations in a 128~atom, 3$\times$3$\times$3 BCC W supercell, using DP-HYB, DP-SE2 and GAP-2, \XY{CL and a MEAM potential proposed by Lee et al.\cite{lee2001second}}.
The DP-SE2 model is highly optimized for the GPU architecture~\cite{jia2020pushing}, but such optimization is only partially available for DP-HYB and not available for the GAP-2 model.
Therefore, for an unbiased comparison, we only use one core of an Intel Xeon Platinum 8269CY CPU to benchmark all models, and the results are shown in Table~\ref{tab:Effic}.
The hybridization of the descriptors makes the DP-HYB 1.6 times more expensive than DP-SE2 model.
The DP-HYB model is more than 3 times faster than GAP-2 potential.
Since the size of the database of GAP-2 is smaller than GAP-1, we would expect that the  GAP-1 model is more expensive than the GAP-2 model.
% The more complex formalism of DP-HYB makes it 1.6 times more expensive than DP-SE2.
% But the DP-HYB are still more than 3 times faster than GAP-2 potential.
% Since the size of database of GAP-2 is already significantly smaller than GAP-1.
% We suggest the relative speed of DP potentials over GAP-1 is even higher.
It is noted that with the recently developed model compression technique, the performance of the DP-SE2 model is further improved up to a factor of 10 with respect to the highly optimized GPU code~\cite{lu2021dp}, which makes the efficiency of the DP-SE2 model comparable to the MEAM potentials.
The efficiency of the DP-HYB model will benefit from the same model compression techniques in the near future, and be made more suitable for large-scale atomistic simulations.
% Moreover, with the recently developed model contraction technique\cite{lu202186,lu2021dp,jia2020pushing}, the DP models are expected to achieve tens of times of boost of efficiency, which makes the efficiencies of DP models comparable to MEAM potentials.
% This enables the DP models suitable to large scale atomistic simulations.

\begin{center}
\begin{table}
\caption{The computational efficiency of the CL, MEAM, DP-SE2, DP-HYB and GAP-2 models. 
The total wall clock time are recorded for 1$\times10^5$ times steps MD simulations of a system consisting 128 atoms.
Only one core of an Intel(R) Xeon(R) Platinum 8269CY CPU @ 2.50GHz is used in all test cases.
% The testing cpu is a single core of Intel(R) Xeon(R) Platinum 8269CY CPU @ 2.50GHz
}\label{tab:Effic}

\begin{threeparttable}
\begin{tabular}{c|ccccc}

\hline
\hline
Potential & CL & MEAM\cite{lee2001second} & DP-SE2 & DP-HYB & GAP-2\\
\hline
Time(ms/step/atom) & 0.002& 0.009 & 0.657 & 1.475 & 4.584\\
\hline
\hline
\end{tabular}
\end{threeparttable}
\end{table}
\end{center}

\section{Conclusive remarks}\label{sec:conclusion}

In this work, a highly generalizable \recheck{(in the sense of in-distribution)} W \recheck{deep neural-network} potential is developed for the purpose of reliable atomistic simulations of property degradation of tungsten that serves under complex multi-physics working condition in the fusion facilities.
This potential utilizes a newly designed descriptor: DP-HYB, which consists of a three-body-embedding descriptor and the original two-body-embedding descriptor.
The representability of the DP-HYB model is significantly enhanced due to the explicit consideration of bond-angle contribution in the embedding matrix.
With the excellent fitting ability of the deep neural networks, the DP-HYB model is fitted to over 40000 configurations generated by the concurrent learning  scheme DP-GEN.
% The major advantage of the DP-HYB model over the empirical formalism is the enhanced representability and ability of generalization.
% We demonstrate that the DP-HYB model is highly generalizable and may accurately describe a broad range of defect structures.

According to our benchmark results, the three-body embedding formalism enables the W DP-HYB potential to achieve the accurate description of a wide range of properties including bulk, elastic, defective properties such as formation energies of free surfaces and grain boundaries, GSF energies, formation energies of prismatic loops, screw dislocation core structure, Peierls barrier and migration trajectory.
{Only the elastically deformed lattice structures, DP-GEN explored bulk and free surfaces, SIA dumbbell, mono-vacancy and GSF structures are explicitly included in the training dataset}, so the correct prediction of the other properties is attributed to the generalizability of the DP-HYB model.
It is noted that the correct prediction of all the benchmarked properties with the same model is by far challenging for the empirical and ML potential models, including the original DP model that only uses the two-body embedding.
Therefore, the DP-HYB is believed a good candidate potential for revealing the underlying physics of mechanical property degradation of W serving under the multi-physics environment.

Although the representability and generalizability  of the DP-HYB model are satisfactory, it is unlikely that the current version can be used to simulate the procedures like the early stage of primary irradiation damage formation,
because \recheck{in the present work the DFT calculation settings are not suitable for very short-range atomistic interactions,}
%\recheck{the training data do not have the configurations with very short inter-atomic distance},
so will fail in predicting the dynamics of atomic collisions with high speed.
% Although the DP-HYB model have a good prediction of SIAs and vacancies, directly using to study the early-stage formation of frenkel-pairs in collision cascade could lead to large error.
To improve the performance in this specific field, coupling with the short-range repulsion models, in a way like  the DP-ZBL\cite{wang2019deep} model does, is suggested.
GAP-2 is suitable for simulating early-stage primary radiation damage formation processes, because it is specially fitted to the short-range repulsive interaction between W.
% The short-range interaction can be considered in the DP models via the refinement procedure using the enlarged training datasets, thus will be one of our next goals to further develop the W DP model.
The extension of the DP-HYB model to the simulation of primary irradiation damage is beyond the scope of the current work, and will be investigated in future works.

In addition, due to previous successes of the DP model in binary and ternary alloys, our future goals include investigating the defect behavior under the influence of alloying/impurity elements.
For example, the effect of Rhenium on the stacking fault energy and screw dislocation core structures, the effect of intersitial impurities on the formation energies of screw dislocation kink-pairs and the grain boundary segregation, the formation of hydrogen blisters and precipitates.
Atomistic simulations of these critical behaviors can be informative to the design of alloys that serves in the multi-physics environment.
Besides W, we believe that DP-HYB formalism can also have wide prospects of application of other engineering materials and alloys that serve under complex and harsh conditions.

\section{Data Availability}
\XY{
The DP models, the training dataset and the scripts used to produce the results are available at \url{https://doi.org/10.5281/zenodo.6466996}. 
The DP models and the training data are also available at  \url{https://dplibrary.deepmd.net/\#/project_details?project_id=202204.002}.
}

\section*{Acknowledgement}
The work is supported by the National Science Foundation of China under Grant No.11871110 and 12122103.

% \begin{thebibliography}{100}
% \end{thebibliography}
\bibliography{ref}
\bibliographystyle{unsrt}

\end{document}

\title{
Supplementary Material for\\ 
A Tungsten Deep Neural-Network Potential for Simulating Mechanical Property Degradation Under Fusion Service Environment
% A Tungsten Deep Potential with High Accuracy and transferability based on a Newly Designed Three-body Embedding Formalism
}
\author{Xiaoyang Wang}
\affiliation{Laboratory of Computational Physics,
  Institute of Applied Physics and Computational Mathematics, Huayuan Road 6, Beijing, P. R.~China}
\author{Yinan Wang}
\affiliation{School of Mathematical Sciences, Peking University, No.5 Yiheyuan Road Haidian District, Beijing, P.R.China 100871}
\author{Linfeng Zhang}
\affiliation{DP Technology, Beijing, 100080}
\affiliation{AI for Science Institute, Beijing, 100084}
\author{Fu-Zhi Dai}
\affiliation{DP Technology, Beijing, 100080}
\affiliation{AI for Science Institute, Beijing, 100084}
\author{Han Wang}
\email{wang\_han@iapcm.ac.cn}
\affiliation{Laboratory of Computational Physics, Institute of Applied Physics and Computational Mathematics, Fenghao East Road 2, Beijing 100094, P.R.~China}
\affiliation{HEDPS, CAPT, College of Engineering, Peking University, Beijing 100871, P.R.~China}
\maketitle

\section{The deep potential model}
The deep potential (DP) model assumes the total energy of a system, denoted by $E$, is the summation of the energy contributions of each atom in the system, denoted by $E_i$, i.e.,
\begin{align}
    E=\sum_{i=1}^N E_i
\end{align}
where $N$ is the number of atoms in the system.
In metallic materials systems, we assume the energy contribution $E_i$ depends on the local environment of the atom. 
To model $E_i$, we define the local environment matrix $\mathcal{R}_i$ of a central atom $i$ being the collection of the relative positions of all its neighbors $j$ within a given cutoff radius $r_c$.
The $j$-th line of the environment matrix is made up by the relative position to the $j$-th neighbor, i.e.
\begin{equation}\label{eqn:envmat}
\{\mathcal{R}_i\}_{j,\cdot}=
s(r_{ij})\times \Big(\frac{x_{ij}}{r_{ij}},\frac{y_{ij}}{r_{ij}},\frac{z_{ij}}{r_{ij}}\Big),
\end{equation}
where $\bm r_{ij}=\bm r_j- \bm r_i$ with $\bm r_i$ being the position of atom $i$, $(x_{ij}, y_{ij}, z_{ij})$ denotes the three Cartesian coordinates of the vector $\bm r_{ij}$, and $r_{ij}=|\bm r _{ij}|$ stands for the distance between the neighbor $j$ and the central atom $i$.
The term $s(r_{ij})$ in Eq.~\eqref{eqn:envmat} is defined as $s(r_{ij})=f_c(r_{ij})/r_{ij}$ with $f_c$ being a switching function smoothly varies from 1 to 0 at the cutoff distance. 
One possible construction of the switching function is
\begin{align}
    f_c (r) = \left \{
    \begin{aligned}
    &1 && r < r_{cs} \\
    &u^3 (-6u^2 + 15 u - 10 ) + 1 && r_{cs} \leq r < r_c\\
    &0 && r_c \leq r
    \end{aligned}
    \right., \quad u = \frac{r-r_{cs}}{r_c - r_{cs}}. \label{equ21} 
\end{align}
By this definition, $f_c$ smoothly decays from 1 to 0 in the range $r_{cs} \leq r \leq r_c$. It can be shown that the second order derivative of $f_c$ is continuous.
The matrix $\mathcal{R}_i$ has $N_m$ lines, where $N_m$ is the maximal possible number of neighbors of any atom in the system. 
If the actual number of neighbors of atom $i$ is smaller than $N_m$, the rest places of the environment matrix are filled by zeros.

% i.e.~$\mathcal{R}_i=\{\bm r_{ij}\ ,j\in L(i,r_c) \}$, where $\bm r_{ij}=\bm r_j- \bm r_i$, and $L(i,r_c)$ is the list of neighbors of central atom $i$ within the cutoff radius $r_c$. 
% \begin{equation}
% \{\mathcal{R}_i\}_j=\Big(\frac{x_{ij}}{r_{ij}},\frac{y_{ij}}{r_{ij}},\frac{z_{ij}}{r_{ij}}\Big)
% \end{equation}
% $r_{ij}=|\bm r _{ij}|$ is the distance between the neighbor $j$ and central atom $i$, and $(x_{ij},y_{ij},z_{ij})$ are the Cartesian coordinate of the relative position $\bm r _{ij}$.

In the DP models, the local environment matrix is first mapped onto a descriptor $\mathcal{D}$, which preserves rotation, translation and permutation symmetries, and then mapped onto $E_i$ via a \textbf{fitting net} $\mathcal{F}$.
\begin{equation}
E_i=\mathcal{F}(\mathcal{D}(\mathcal{R}_i))
\end{equation}
In the following sections, we introduce in detail the construction of the descriptor and the fitting net. 

\section{Descriptors}
In this section we will introduce two
types of descriptors, the two-body embedding descriptor that considers the embedding of only distances between atoms, 
and the three-body embedding descriptor that considers the embedding of the inner product of the relative positions of any two neighbors. 
The two-body embedding descriptor was originally introduced in Ref.~\cite{zhang2018end} as the descriptor for the smooth edition of the Deep Potential. 
The three-body embedding descriptor is proposed by this work. 

\subsection{Two-body embedding descriptor}

In the construction of the two-body embedding descriptor $\mathcal{D}_i^{(2)}$, we firstly introduce the \textit{generalized environment matrix} $\mathcal{\Tilde{R}}_i$. 
This matrix has the same number of lines as the environment matrix, but has one more column made up of $s(r_{ij})$. 
The $j$-th row of $\mathcal{\Tilde{R}}_i$ is defined as 
\begin{equation}
\mathcal{\{\Tilde{R}}_i\}_{j, \cdot}=s(r_{ij})\times\Big(1,\frac{x_{ij}}{r_{ij}},\frac{y_{ij}}{r_{ij}},\frac{z_{ij}}{r_{ij}}\Big)
\end{equation}
Similar to the environment matrix $\mathcal R_i$, if the number of neighbors of $i$ is smaller than $N_m$, the empty positions of the generalized environment matrix $\tilde{\mathcal R}_i$ are filled with zeros.

We then introduce the \textit{two-body embedding matrix} $\mathcal G_i^{(2)}$ that involves the embedding of two-atom distances. 
The embedding matrix $\mathcal G_i^{(2)}$ has $N_m$ lines and $M_2$ columns. The $j$-th line is defined as 
\begin{equation}\label{eqn:embdmat2}
(\mathcal{G}_i^{(2)})_{j,\cdot} = \Big( G_1^{(2)}(s(r_{ij})), 
\cdots
G_{M_2}^{(2)}(s(r_{ij}))
\Big),
\end{equation}
where ${G}^{(2)}$, the \textit{two-body embedding net}, being a full connected deep neural network, maps the scaler $s(r_{ij})$ onto $M_2$ outputs. 
The embedding net ${G}^{(2)}$ has $m+1$ layers, and can be mathematically written as 
\begin{equation}\label{eqn:embd-2-net}
G^{(2)}(x)=\mathcal{L}^e_{m}\circ \mathcal{L}^e_{m-1}\circ \cdots \circ \mathcal{L}^e_1\circ\mathcal{L}^e_0(x),
\end{equation}
where $\circ$ denotes the function composition.
The first hidden layer $\mathcal{L}^e_0$ takes a scalar as input and outputs a vector of size $s_0$. It is defined by
\begin{equation}
\label{hidden_layer1_embedding}
\mathcal{L}^e_0(x)=\tanh(x\cdot W^e_0+b^e_0),
\end{equation}
where $W^e_0 \in \mathbb R^{s_0}$ are the weights, represented by a vector of size $s_0$, $b^e_0 \in \mathbb R^{s_0} $ denote the biases and the activation function $\tanh$ applies to the vector $x\cdot W_0^e + b_0^e$ in a component-wise way.
Other hidden layers are expressed as 
\begin{equation}
\label{hidden_layern_embedding}
\mathcal{L}^e_k(x)=(x,x)+ \tanh(x\cdot W^e_k+b^e_k),\quad 1<k\leq m
\end{equation}
where $(x,x)$ denotes the concatenation of two input vectors $x$. 
The weights are $W_k^e \in \mathbb R^{s_{k-1}\times s_k}$, and the biases are $b_k^e \in \mathbb R^{s_k}$.
We let the output size of the $k$-th hidden layer be twice of the input size, i.e. $s_{k}=2s_{k-1}$.
The output size of the final layer, $s_m$, is equal to $M_2$, which is the same as the number of columns of the embedding matrix $\mathcal{G}^{(2)}_i$.
The parameters in the embedding net, $\{W_k^e, b_k^e\}_{k=0}^m$ will be trained together with the fitting net in an end-to-end way. 
The training of the DP model will be explained in Sect.~\ref{sec:training}.

% In $\mathcal{D}^{(2)}$, the embedding net is evaluated for each pair of neighbors $(i,j)$, thus the computational complexity of $\mathcal{D}^{(2)}$ is $\mathcal{O}(N\times N_m)$, which is in proportion with the total number of atoms $N$, and maximum number of neighbors $N_m$ .

\begin{figure}
    \centering
    \includegraphics[width=0.98\textwidth]{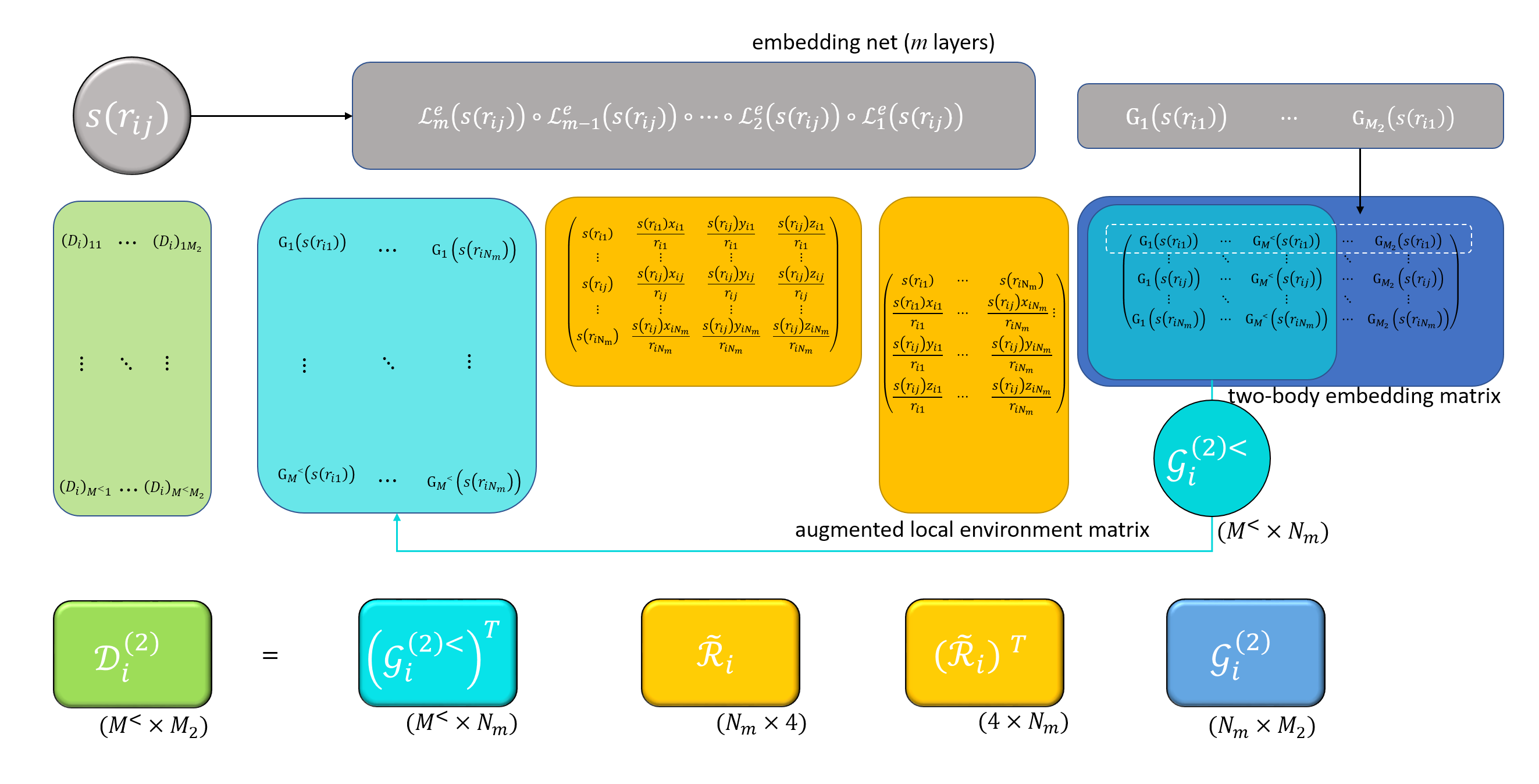}
    \caption{Schematics of the two-body descriptor.}
    \label{fig:descriptor2}
\end{figure}

With the generalized environment and the two-body embedding matrices in hand, we are ready to construct the two-body embedding descriptor $\mathcal{D}_i^{(2)}$:
\begin{align}\label{eqn:descriptor2}
   \mathcal D^{(2)}_i = 
    \frac{1}{N_m^2}
    ( \mathcal G_i^{(2),<} )^T 
    \tilde{\mathcal R}_i 
    (\tilde{\mathcal R}_i)^T
    \mathcal G_i^{(2)},
\end{align}
where the superscript $T$ denotes the matrix transpose.
The superscript $<$ on $\mathcal G_i^{(2),<}$ means that $\mathcal G_i^{(2),<}$ is a sub-matrix of $\mathcal G_i^{(2)}$, taking the first $M^<$ columns of the latter matrix.
The descriptor $ \mathcal D_i^{(2)}$ is a matrix of shape $M^< \times M_2$, and is reshaped in to a vector before it is passed to the fitting net. 
The construction of the two-body embedding descriptor $\mathcal{D}_i^{(2)}$ is schematically illustrated in Fig.\ref{fig:descriptor2}.

\subsection{Symmetries of the two-body embedding descriptor}

The two-body embedding descriptor $\mathcal D_i^{(2)}$ preserves the translation symmetry, because all elements of the generalized environment  and the embedding matrices are functions of relative positions between atom $i$ and its neighbors. 

The two-body embedding matrix $\mathcal G_i^{(2)}$ is invariant to rotation transform, because the input of the embedding net is the distance between $i$ and its neighbors, which is invariant to rotation. 
The matrix $\tilde{\mathcal R}_i (\tilde{\mathcal R}_i)^T$ is over-complete, with the $jk$-th element defined by
\begin{align}
   \{ \tilde{\mathcal R}_i (\tilde{\mathcal R}_i)^T \}_{jk} 
   = 
   s(r_{ij})s(r_{ik}) \times 
   \Big(
   1 + \frac{\bm r_{ij}\cdot \bm r_{ik}}{r_{ij}r_{ik}}
   \Big).
\end{align}
The $jk$-th element is invariant under rotational transform, because the inner product  $\bm r_{ij}\cdot \bm r_{ik}$ and the distances $r_{ij}$, $r_{ik}$ are invariant under rotation.
Therefore the two-body embedding descriptor $\mathcal D_i^{(2)}$ preserves the rotational symmetry.

The $\alpha\beta$-th element of the matrix product $(\tilde{\mathcal R}_i)^T\mathcal G_i^{(2)}$ is written as
\begin{align}\label{eq:rtg-element}
    \{(\tilde{\mathcal R}_i)^T\mathcal G_i^{(2)}\}_{\alpha,\beta} 
    = 
    \sum_{j=1}^{N_m}
    \{\tilde{\mathcal R}_i\}_{\alpha,j}
    \{\mathcal G_i^{(2)}\}_{j,\beta},
\end{align}
where $\{\tilde{\mathcal R}_i\}_{\alpha j}$ and $\{\mathcal G_i^{(2)}\}_{j\beta}$ are the $\alpha j$-th and $j \beta$-th element of the generalized environment and the embedding matrices, respectively. 
The summation in Eq.~\eqref{eq:rtg-element} ensures that the product $(\tilde{\mathcal R}_i)^T\mathcal G_i^{(2)}$ is invariant under any change in the order of neighbors of the same chemical species, thus preserves the permutational symmetry. 
Similar argument applies to the product $( \mathcal G_i^{(2),<} )^T \tilde{\mathcal R}_i $, thus the two-body embedding descriptor preserves the permutational symmetry.

\subsection{Three-body embedding descriptor}

\begin{figure}
    \centering
    \includegraphics[width=0.98\textwidth]{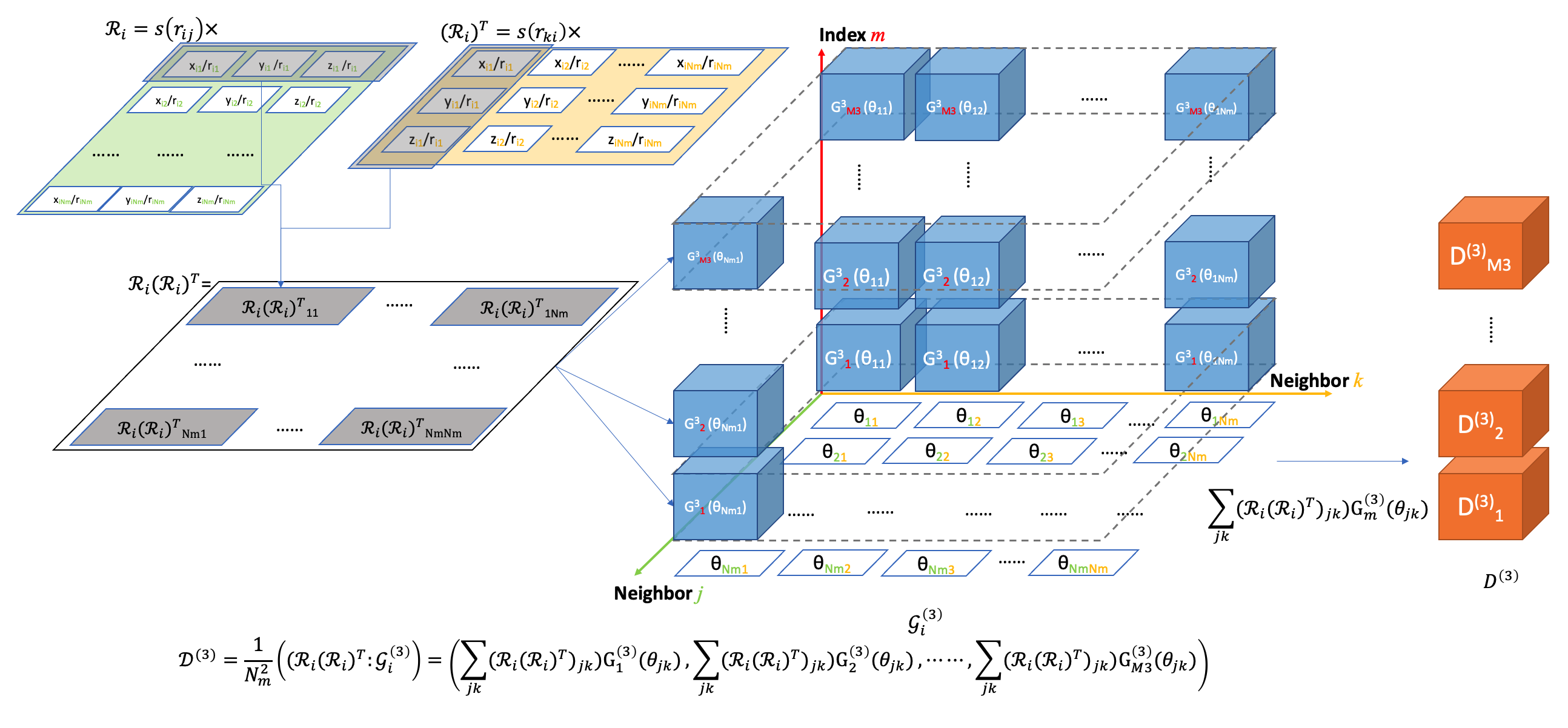}
    \caption{Schematics of the three-body descriptor.}
    \label{fig:descriptor3}
\end{figure}

The three-body embedding descriptor $\mathcal{D}_i^{(3)}$ is distinguished from the two-body embedding descriptor $\mathcal{D}^{(2)}$ by the embedding matrix. 
We firstly introduce a short-hand notation $(\theta_i)_{jk}$ for the elements of the product of the environment matrix with its transpose. By Eq.~\eqref{eqn:envmat}, we have
\begin{align}\label{eqn:thetaijk}
    (\theta_i)_{jk} \equiv 
    \{\mathcal R_i (\mathcal R_i)^T\}_{jk}
    =
    s(r_{ij})s(r_{ik})\frac{\bm r_{ij}\cdot\bm r_{ik}}{r_{ij} r_{ik}}.
\end{align}
For any atom $i$ we define the three-body embedding tensor $\mathcal{G}_i^{(3)}$, which is an order-3 tensor, as 
\begin{equation}\label{eqn:embdmat3}
(\mathcal{G}_i^{(3)})_{jk,\cdot} = 
\Big(
G_1^{(3)} \big( (\theta_i)_{jk} \big),
\cdots,
G_{M_3}^{(3)} \big( (\theta_i)_{jk} \big)
\Big),
\end{equation}
where $G^{(3)}$, called the \emph{three-body embedding net}, maps the scalar $(\theta_i)_{jk}$ to a vector of dimension $M_3$.
It is represented by a full connected feed forward deep neural network, which has the same architecture as the two-body embedding net defined by Eqs.~\eqref{eqn:embd-2-net}--\eqref{hidden_layern_embedding}.
We denote the number of layers of the three-body embedding net by $m_t+1$, and denote the trainable parameters of it by $\{W^t_k, b_k^t\}_{k=0}^{m_t}$. 
The first two indices, i.e.~$j$ and $k$, of the embedding tensor $\mathcal{G}_i^{(3)}$ go from 1 to $N_m$.

Based on the environment matrix and the three-body embedding matrix, we propose the following form of the three-body embedding descriptor $\mathcal{D}_i^{(3)}$ 
\begin{equation}\label{eqn:descriptor3}
\mathcal{D}_i^{(3)} = \frac{1}{N_m^2}\theta_i :\mathcal{G}_i^{(3)} 
= \frac{1}{N_m^2} \sum_{jk=1}^{N_m} (\theta_i)_{jk} (\mathcal{G}_i^{(3)})_{jk}, 
\end{equation}
where $:$ denotes the double contraction operation.  The construction of the three-body descriptor $\mathcal{D}_i^{(3)}$ is schematically explained in Fig.\ref{fig:descriptor3}.
It can be understood as multiplying each element in $\mathcal{G}_i^{(3)}$ with the element having the same $(j,k)$ subscript in the $N_m \times N_m$ matrix $\theta_i$, then sum them all together as one element in $\mathcal{D}_i^{(3)}$. 
The descriptor is a vector of dimension $M_3$.

\subsection{Symmetries of the three-body embedding descriptor}

The three body embedding descriptor is invariant under the translation of the positions of atoms, because the environment matrix $\mathcal R_i$ is defined as a function of relative positions between atom $i$ and its neighbors. 

By Eq.~\eqref{eqn:thetaijk}, $(\theta_i)_{ij}$ is invariant with rotational transform, because we have dot product of $\bm r_{ij}$ and $\bm r_{ik}$ and their lengths on the right-hand-side of the definition, which are all invariant under rotation. 
This leads to the fact that the three-body embedding tensor $\mathcal G_i^{(3)}$, as a function of $(\theta_i)_{ij}$, is also invariant with rotational transforms. 
Thus the descriptor $\mathcal{D}_i^{(3)}$ preserves the rotational symmetry. 

The summation over neighbor indices $j$ and $k$ in Eq.~\eqref{eqn:descriptor3} ensures that the descriptor $\mathcal{D}_i^{(3)}$ in invariant with any change in the order of neighbors of the same chemical species, thus the permutational symmetry is preserved.

% ${\mathcal R}_i ({\mathcal R}_i)^T$ preserves the translation and rotation symmetry, and the double contraction operation makes $\mathcal{D}^{(3)}$ invariant with respect to the permutation of neighbors $j,k$. Thus the three-body descriptor also preserves the rotation, translation and permutation symmetry. 

\subsection{Hybridization of descriptors and the computational complexity}

In our DP-HYB model, the descriptor is the hybridization of the two-body embedding descriptor $\mathcal{D}_i^{(2)}$ and our newly proposed three body embedding descriptor $\mathcal{D}_i^{(3)}$, i.e.
\begin{equation}
\mathcal{D}_i=(\mathcal{D}_i^{(2)},\mathcal{D}_i^{(3)}),
\end{equation}
where the notation $(\cdot, \cdot)$ means that both descriptors $\mathcal{D}^{(2)}$ and $\mathcal{D}^{(3)}$ are treated as vectors, and the two vectors are concatenated to form a new vector. 
Thus the hybrid descriptor $\mathcal{D}_i$
has a total number of $M^<\times M_2+M_3$ outputs.
The parameters needed to be determined in the training are $\{W_k^e, b_k^e\}_{k=0}^m$, $\{W_k^t, b_k^t\}_{k=0}^{m_t}$.

% Instead of using only the function of neighbor distances $s(r_{ij})$ as input in $\mathcal{D}^{(2)}$, the embedding matrix of $\mathcal{D}^{(3)}$, namely $\mathcal{G}^{(3)}$, also takes the bond-angle of each pair of neighbors, ${\theta_i}_{jk} $ as inputs. Thus the  $\mathcal{D}^{(3)}$ explicitly considers the contributions of bond-angles of each pair of neighbors. For the central atom $i$ and each pair of neighboring atoms $j,k$ , the embedding matrix is presented in form:

The computational cost of the environment $\mathcal R_i$ and the two-body embedding $\mathcal G^{(2)}_i$ matrices are of order $\mathcal{O}( N_m)$. 
The cost of matrix multiplications in Eq.~\eqref{eqn:descriptor2} is $\mathcal O(N_m)$. 
If we denote the number of atoms in the system by $N$, then the total computational cost of two-body descriptors of all atoms is  $\mathcal{O}(N\times N_m)$, which is in proportion with the total number of atoms $N$, and maximum number of neighbors $N_m$.
The computational cost of the three-body embedding tensor $\mathcal G^{(3)}_i$ is $\mathcal O(N_m^2)$. 
The cost of the double contraction in Eq.~\eqref{eqn:descriptor3} is also $\mathcal O (N_m^2)$.
The computational cost of the three-body embedding descriptors of all atoms is $\mathcal{O}(N\times N_m^2)$.
Considering the computational cost, the cut-off radius used to calculate the three-body embedding descriptor is usually chosen to be smaller than that used to calculate the two-body embedding descriptor.

% In $\mathcal{D}^{(3)}$, the embedding net is evaluated for each of pairs of neighbors $(k,j)$ of atom $i$, thus the computational complexity of $\mathcal{D}^{(3)}$ is $\mathcal{O}(N\times N_m^2)$, which is tremendously greater than that of $\mathcal{D}^{(2)}$. Considering the computational cost, the cut-off radius $r_c$ and size of the embedding net $M_3$ should be significantly less than that of $\mathcal{G}^{(2)}$. $\mathcal{D}^{(3)}$ has $M_3$ outputs. Thus the hybrid descriptor $\mathcal{D}$
% has a total number of $M^<\times M_2+M_3$ outputs.

\section{Fitting Net}
The fitting net maps the descriptor $\mathcal D_i$ to the  energy contribution $E_i$ of each atom $i$. 
Fitting net is a fully connected deep neural network with skip-connections, and contains $l$ hidden layers. It is written as
\begin{equation}
\mathcal F(x)=\mathcal{L}^f_{l}\circ \cdots \circ \mathcal{L}^f_1\circ\mathcal{L}^f_0(x).
\end{equation}
The layers of the fitting net are defined as 
\begin{align}
    & \mathcal{L}^f_0(x)=\tanh(x\cdot W^f_0+b^f_0) \\ \label{eqn:fit-net-2}
    & \mathcal{L}^f_k(x)=x+\tanh(x\cdot W^f_k+b^f_k), \quad 1\leq k < l \\
    & \mathcal{L}^f_l(x) = x\cdot W^f_l+b^f_l
\end{align}
In the first hidden layer $\mathcal{L}^f_0$, the weights $W_0^f$ and bias $b_0^f$ are of size $(M^<\times M_2 + M_3) \times M_F$ and $M_F$, respectively. 
In the  layers $1 \leq k < l$, the input and output are vectors of the same length $M_F$, thus a skip connect is setup, see Eq.~\eqref{eqn:fit-net-2}. 
The weights and biases are of size $M_F \times M_F$ and $M_F$, respectively.
The output layer $\mathcal{L}^f_l$ is a linear transform that maps vector of length $M_F$ to a scalar. The weights $W_l^f$ form a vector of size $M_F$, and the bias is a scalar. 
All the parameters $\{W^f_k, b^f_k\}_{k=0}^l$ are optimized together with the parameters in the descriptor during the training.

% All the hidden layers of the fitting net has the same size $M_F$. Within each hidden layer, a skip connection between the input and output is used.
% \begin{equation}
% \label{hidden_layers_fitting}
% \mathcal{L}^f_k(x)=x+\tanh(x\cdot W^f_k+b^f_k)
% \end{equation}
% $W^f_k,b^f_k$ are the weights and biases of each hidden layers in the fitting net.
% The activation function $\tanh$ is used in the fitting net.
% In the input layer, the weights $W_0^f$ and bias $b_0^f$ are of size $(M^<\times M_2 + M_3) \times M_F$ and $M_f$, respectively. 
% In the hidden layers, $1 \leq k < l$, the weights and biases are of size $M_F \times M_F$ and $M_F$, respectively.
% In the output layer, the weights $W_l^f$ form a vector of size $M_F$, and the bias is a scalar. 
% All the parameters $\{W^f_k, b^f_k\}_{k=0}^l$ are optimized together with the parameters in the descriptor during the training.

\section{Training}\label{sec:training}

In the DP model, the system energy is the summation of atomic energy contributions, 
\begin{align}
    E = \sum_i E_i = \sum_i \mathcal F(\mathcal D_i) = \sum_i \mathcal F(\mathcal D(\mathcal R_i) ).
\end{align}
Since the descriptor is a function of the environment matrix, we denote $\mathcal D_i = \mathcal D(\mathcal R_i)$.
The forces $F_i$ of an atom $i$ can be calculated by
\begin{equation}\label{eqn:f}
F_i=-\nabla _{r_i} E
\end{equation}
The virial tensor of a system is defined as:
\begin{equation}\label{eqn:v}
\Xi_{\alpha\beta}=-\frac{\partial E}{\partial{h_{\gamma\alpha}}}{h_{\gamma\alpha}}
\end{equation}
where ${h_{\alpha\beta}}$ is the $\beta$ th component of $\alpha$ th basis vector of the simulation cell.
% The derivatives of the energy in Eqs.~\eqref{eqn:f} and \eqref{eqn:v} are can be easily implemented by the auto-differentiation in deep learning frameworks like the Tensorflow

During the training processes, the weights and biases of the embedding nets ($\{W_k^e, b_k^e\}_{k=0}^m$, $\{W_k^t, b_k^t\}_{k=0}^{m_t}$) and those of the  fitting net ($\{W^f_k, b^f_k\}_{k=0}^l$) are trained to minimize the loss function 
\begin{align}\label{eq:loss}
    \mathcal L &= \frac{1}{\vert \mathcal B\vert} 
    \sum_{k\in \mathcal B}
    \Big(
     p_\epsilon \frac 1N \vert \hat E^k - E^k\vert^2 + 
     p_f \frac 1{3N} \sum_{i\alpha} \vert \hat F_{i\alpha}^k - F_{i\alpha}^k \vert^2 +
     p_\xi \frac 1{9N} \sum_{i\alpha} \vert \hat \Xi_{\alpha\beta}^k - \Xi_{\alpha\beta}^k \vert^2
    \Big), 
\end{align}
% \begin{equation}
%     L(p_\epsilon,p_f,p_\xi)=\frac{1}{|\mathcal{B}|}\sum_{l\in\mathcal{B}}p_\epsilon |E_l-E_l^w|^2+p_f |F_l-F_l^w|^2+p_\xi ||\Xi_l-\Xi_l^w||^2,
% \end{equation}
which measures the difference between the DFT energy $\hat E^k$, forces $\hat F_{i\alpha}^k$ and the virial tensor $\hat \Xi_{\alpha\beta}^k$, and those predicted by the model.
In this work, the Adam stochastic gradient descent optimizer~\cite{Kingma2015adam} is used in the training.
% obtained via a training process.
% A loss function is minimized with the Adam stochastic gradient descent method, and the loss function form is defined as:
% \begin{equation}
%     L(p_\epsilon,p_f,p_\xi)=\frac{1}{|\mathcal{B}|}\sum_{l\in\mathcal{B}}p_\epsilon |E_l-E_l^w|^2+p_f |F_l-F_l^w|^2+p_\xi ||\Xi_l-\Xi_l^w||^2
% \end{equation}
In Eq.~\eqref{eq:loss}, $\mathcal{B}$ is a mini-batch of datasets, and $|\mathcal{B}|$ denotes the batch size.
The superscript $k$ denotes the index of the training data in the mini-batch.
Each training datum contains the configuration of the system (including the coordinates of atoms, the box basis vectors and the element types), and the corresponding labels (total energy, forces on each atoms, and the virial tensor), which are obtained by a DFT calculation. 
% Instead of directly fit to material properties(such as elastic constant), deep-potential model are fitted to the DFT-calculated energies $E_l$, forces $F_l$ and virial tensors $\Xi_l$ of configurations  $l$ in $\mathcal{B}$. 
Prefactors $(p_\epsilon,p_f,p_\xi)$ are a set of hyper-parameters determining the relative importance of the energy, forces and virial tensor during the training. 
The prefactors are gradually adjusted according to the {learning rate} $r_l(t)$, which exponentially decays with the training step $t$:
\begin{align}\label{eq:learning rate}
r_l(t)=r_l^0k_d^{t/t_d},
\end{align}
where $r_l^0$ is the learning rate at the beginning, $t_d$ denotes the typical timescale of learning rate decaying and $k_d$ denotes the decay rate. 
The prefactors vary with the learning rate in the following way
\begin{align}\label{eq:pref_decay}
p_{\alpha}(t)=p_{\alpha}^{limit}[1-\frac{r_l(t)}{r_l^{0}}]+p_{\alpha}^{start}[\frac{r_l(t)}{r_l^{0}}], 
\quad \alpha = \epsilon,\ f,\ \textrm{or}\ \xi,
\end{align}
$p_{\alpha}(t)$ is either of the three pre-factors $(p_\epsilon,p_f,p_\xi)$ at training step $t$. $p_{\alpha}^{start}$ and $p_{\alpha}^{limit}$ are the pre-factors at the beginning and at an infinitely small learning rate.
The prefactors change linearly from the $p_{\alpha}^{start}$ to $p_{\alpha}^{limit}$ during the training process. 
In practice, a relatively larger force prefactor at the beginning and balanced prefactors at the end of the training can make the best use of the training datasets and achieve relatively good accuracy. 

\section{Hyper-Parameters}\label{sec:hyper-parameter}
% During the training processes in each iterations of the concurrent learning scheme DP-GEN, the list of  hyper-parameters are shown in table S1.

The training parameters used during the training step of the DP-GEN scheme and those used for training the productive DP models are summarized in Tab.~\ref{tab:hyper}.

\begin{table}[]
    
\label{tab:hyper}
    \centering
    \caption{Hyper-parameters used during the training in the DP-GEN iterations and the training of the productive models. 
    The parameters used in the first and second refinements are listed in parenthesis respectively.
    The superscript (2) and (3) denotes the hyper-parameter used in two-body embedding descriptors $\mathcal{D}^{(2)}$ and three-body embedding descriptors $\mathcal{D}^{(3)}$, respectively.
    }
    \begin{tabular}{c|c|c|c}
    \hline
    \hline
         Hyper-parameter  & DP-GEN Iterations & Productive Model DP-SE2 & Productive Model DP-HYB\\
         \hline
            $r_c^{(2)}$& 6.0{\AA}&6.0{\AA}&6.0{\AA}\\
            $r_{cs}^{(2)}$& 2.0{\AA}&2.0{\AA}&2.0{\AA}\\
            $N_m^{(2)}$& 100&100&100\\
            layers of $\mathcal{G}^{(2)}$ &20,40,80&20,40,80&20,40,80\\
            $M_2$& 80& 80& 80\\
            $M^<$& 16& 16& 16\\
            \hline
            $r_c^{(3)}$& -&-&4.0{\AA}\\
            $r_{cs}^{(3)}$& -&-&2.0{\AA}\\
            $N_m^{(3)}$& -&-&32\\
            layers of $\mathcal{G}^{(3)}$ &-&-&4,8,16\\
            $M_3$& - &-&16\\
            \hline
            $M_F$&240&240&240\\
            $r_l^{0}$& 0.001&0.001(0.0001,0.0001)&0.001(0.0001,0.0001) \\
            $k_d$& 0.95 &0.95&0.95\\
            $t_d$& $2\times 10^3$ &$120\times 10^3$($5\times 10^3$,$20\times 10^3$ ) &$120\times 10^3$($5\times 10^3$, $20\times 10^3$)\\
            $p^{start}_{\epsilon}$ & 0.02&0.02(1.0, 1.0)&0.02(1.0, 1.0)\\
            $p^{limit}_{\epsilon}$ & 1.0&1.0(1.0, 1.0)&1.0(1.0, 1.0)\\
            $p^{start}_{f}$ & 1000.0&1000.0(1.0, 1.0)&1000.0(1.0, 1.0)\\
            $p^{limit}_{f}$ & 1.0&1.0(1.0, 1.0)&1.0(1.0, 1.0)\\
            $p^{start}_{\xi}$ & 0.02&0.02(0.9, 0.9)&0.02(0.9, 0.9)\\
            $p^{limit}_{\xi}$ & 1.0&1.0(1.0, 1.0)&1.0(1.0, 1.0)\\
            Activation Function & tanh&tanh&tanh\\
            \hline
            Training Steps & 0.4$\times 10^6$&$24\times 10^6(1\times 10^6,4\times 10^6)$&$24\times 10^6(1\times 10^6,4\times 10^6)$\\

    \hline
    \hline
    \end{tabular}
    \label{tab:hyper}

\end{table}

\bibliography{ref}
\bibliographystyle{unsrt}